\documentclass[prd,aps,twocolumn,preprintnumbers, showpacs, nofootinbib,superscriptaddress,notitlepage]{revtex4-1}
\usepackage{amsmath,graphicx,epsfig,amssymb}
\usepackage[utf8]{inputenc}
\usepackage[usenames]{color}
\usepackage{ulem} 
\usepackage{bigstrut}
\usepackage{slashed}
\usepackage{multirow}
\usepackage{subfigure}
\usepackage{dsfont}
\usepackage{xcolor}

\allowdisplaybreaks

\newcommand{\nn}{\nonumber}

\begin{document}
 
\title{
Calculation of  heavy meson light-cone distribution amplitudes from lattice QCD}

\author{
Xue-Ying Han}
\affiliation{Institute of High Energy Physics, Chinese Academy of Sciences, Beijing 100049, China}
\affiliation{School of Physical Sciences, University of Chinese Academy of Sciences, Beijing 100049, China}

\author{Jun Hua}
\affiliation{Key Laboratory of Atomic and Subatomic Structure and Quantum Control (MOE), Guangdong
Basic Research Center of Excellence for Structure and Fundamental Interactions of Matter,
Institute of Quantum Matter, South China Normal University, Guangzhou 510006, China }
\affiliation{Guangdong-Hong Kong Joint Laboratory of Quantum Matter, Guangdong Provincial Key Laboratory of Nuclear Science, Southern Nuclear Science Computing Center, South China Normal University, Guangzhou 510006, China}

\author{Xiangdong Ji}
\affiliation{Maryland Center for Fundamental Physics, Department of Physics, University of Maryland, 4296 Stadium Dr., College Park, MD 20742, USA}

\author{Cai-Dian L\"u}
\affiliation{Institute of High Energy Physics, Chinese Academy of Sciences, Beijing 100049, China}
\affiliation{School of Physical Sciences, University of Chinese Academy of Sciences, Beijing 100049, China}

\author{Andreas Sch\"afer}
\affiliation{Institut f\"ur Theoretische Physik, University Regensburg
D-93040 Regensburg, Germany}

\author{Yushan Su}
\affiliation{Maryland Center for Fundamental Physics, Department of Physics, University of Maryland, 4296 Stadium Dr., College Park, MD 20742, USA}

\author{Wei Wang}
\email{Corresponding author: wei.wang@sjtu.edu.cn}
\affiliation{Shanghai Key Laboratory for Particle Physics and Cosmology, Key Laboratory for Particle Astrophysics and Cosmology (MOE), School of Physics and Astronomy, Shanghai Jiao Tong University, Shanghai 200240, China}

\affiliation{Southern Center for Nuclear-Science Theory (SCNT), Institute of Modern Physics, Chinese Academy of Sciences, Huizhou 516000, Guangdong Province, China}

\author{Ji Xu}
\affiliation{School of Nuclear Science and Technology, Lanzhou University, Lanzhou 730000, China}
\affiliation{School of Physics and Microelectronics, Zhengzhou University, Zhengzhou, Henan 450001, China}

\author{Yibo Yang}
\affiliation{School of Physical Sciences, University of Chinese Academy of Sciences, Beijing 100049, China}
\affiliation{CAS Key Laboratory of Theoretical Physics, Institute of Theoretical Physics,
Chinese Academy of Sciences, Beijing 100190, China}
\affiliation{School of Fundamental Physics and Mathematical Sciences,
Hangzhou Institute for Advanced Study, UCAS, Hangzhou 310024, China}
\affiliation{International Centre for Theoretical Physics Asia-Pacific, Beijing/Hangzhou, China}

\author{Jian-Hui Zhang}
\affiliation{School of Science and Engineering, The Chinese University of Hong Kong, Shenzhen 518172, China}

\author{Qi-An Zhang }
\email{Corresponding author: zhangqa@buaa.edu.cn}
\affiliation{School of Physics, Beihang University, Beijing 102206, China}

\author{Shuai Zhao}
 \affiliation{School of Science, Tianjin University, Tianjin 300072, China}

\collaboration{Lattice Parton Collaboration}

\begin{abstract}
We  develop an approach for calculating heavy quark effective theory (HQET) light-cone distribution amplitudes (LCDAs) by employing a sequential effective theory methodology. The theoretical foundation of the framework is established, elucidating how the quasi distribution amplitudes (quasi DAs) with three scales can be utilized to compute HQET LCDAs. We  provide theoretical support for this approach by demonstrating the rationale behind devising a hierarchical ordering for the three involved scales, discussing the factorization at each step,  clarifying the underlying reason for obtaining HQET LCDAs in the final phase,  and addressing potential theoretical challenges.   The lattice QCD simulation aspect is explored in detail, and the computations of  quasi DAs are presented.   We employ three  fitting strategies to handle  contributions from excited states and extract the bare matrix elements. For renormalization purposes, we  apply hybrid renormalization schemes at short and long distance separations. To mitigate long-distance perturbations, we perform an extrapolation in $\lambda= z\cdot P^z$ and assess the stability against various parameters.  After two-step matching, our results for HQET LCDAs  are found in agreement with existing model parametrizations.  The potential phenomenological implications of the results are discussed, shedding light on how these findings could impact our understanding of the strong interaction dynamics and physics beyond the standard model. It should be noted, however, that systematic uncertainties have not been accounted for yet.
\end{abstract}

\maketitle


\section{Introduction}

Heavy meson light-cone distribution amplitudes (LCDAs) encapsulate details regarding the likelihood of having a heavy quark and  a light (anti)quark  with specific momentum sharing inside a heavy meson~\cite{Grozin:1996pq}. These LCDAs are defined within the framework of heavy quark effective theory   (HQET)  as:
\begin{align}
\label{eq:HQETLCDA}
 \langle 0|O_{v}^P(t) &|H(p_H)\rangle = i \tilde f_H   m_H n_+\cdot v  \nn\\
& \times \int_0^\infty d\omega e^{i \omega t n_+\cdot v} \varphi^+(\omega;\mu),
\end{align} 
where $\omega$ is the momentum carried by the light quark,  $\tilde f_H$ is the HQET decay constant, and the involved  operators are:  
\begin{align}
O_{v}^P(t) =& \bar q_s(tn_+) \slashed{n}_+\gamma_5W_c[tn_+,0]{h}_v(0). 
\end{align}
Here   the $h_v$ is the effective operator for a heavy quark  moving along the $v$ direction and   $W_c[tn_+,0]$ is a finite-distance Wilson line along the light-like vector $n_+$ ($n_+^2=0$).  $q_s$ represents a light quark field with soft momentnum, and $t$ denotes the separation along the light-cone between $q_s$ and $h_v$. The presence of the Wilson line connecting the soft quark field $q_s$ and $h_v$  ensures gauge invariance in the context of HQET.   Heavy meson LCDAs are needed for the description of exclusive reactions and thus play a crucial role in elucidating the dynamics of the strong force at the boundary between long-range hadronic characteristics and short-range quark-gluon aspects. 

On the theoretical front,  heavy meson LCDAs are crucial in predicting decay widths and other decay characteristics of  heavy bottom mesons, which offer a valuable testing ground for the standard model of particle physics. This also  allows us to explore potential new physics phenomena and enhance our comprehension of the strong and weak nuclear forces.  Within  QCD factorization~\cite{Beneke:1999br,Beneke:2000ry} (see Refs.~\cite{Keum:2000wi,Lu:2000em} for an alternative  scheme based on $k_T$ factorization), once the highly off-shell degrees of freedom have been integrated out, the matrix elements for nonleptonic decays, such as $\overline B^0\to \pi^+\pi^-$, can be expressed as:
\begin{align}\label{fform}
&\langle\pi^+\pi^-|Q_i|\overline B^0(p)\rangle =
f^0_{B\to\pi}(m_\pi^2)\int^1_0 dx\, T^{I}_i(x;\mu)\Phi_\pi(x;\mu)  \nn\\
&~~  +\int^1_0 d\xi dx dy T^{II}_i(\xi,x,y;\mu)\varphi^+(\xi;\mu) \Phi_\pi(x;\mu) \Phi_\pi(y;\mu),
\end{align}
where  $\mu$ is the factorization scale. $Q_i$  represents a four-quark operator, $T_i^I$ and $T_i^{II}$ are the short-distance coefficients, $f^0_{B\to\pi}(q^2)$
denotes a $B\to \pi$ form factor with $q^2$ being the momentum transfer square which is equal to $m_\pi^2$ in the present case. $\Phi_\pi$ and $\varphi^+$ correspond to the pion and heavy $B$ meson LCDAs, respectively. Therefore, it is evident that a thorough understanding of heavy meson LCDAs is essential for accurately predicting decay widths and various other observables.

Though their ultraviolet behavior is calculable   from  QCD perturbation theory~\cite{Lange:2003ff,Lee:2005gza,Kawamura:2008vq,Bell:2013tfa,Feldmann:2014ika,Braun:2019wyx}, the precise shapes of LCDAs  are quite uncertain.   Many model parametrizations are proposed~ \cite{Wang:2015vgv,Beneke:2018wjp,Gao:2021sav} with parameters studied in \cite{Grozin:1996pq,Lee:2005gza,Braun:2003wx,Lan:2019img,Khodjamirian:2020hob,Rahimi:2020zzo,Serna:2020txe},  and are commonly employed in phenomenological investigations.  For instance, recent studies  have utilized these models in the framework of $B$ meson light-cone sum rules (LCSRs) to calculate the form factors for  $B\to K^*$  and $B\to \pi$~\cite{Gao:2019lta,Cui:2022zwm}~\footnote{We thank Yuming Wang for providing the error budget for $B\to \pi$ form factor. }.  The  obtained results at $q^2=0$  are as follows:
\begin{align} \label{eq:B-LCSR_formfacor}
{\cal V}_{B\to K^*}(0) =& 0.359^{+0.141}_{-0.085}\Big|_{\lambda_B}{}^{+0.019}_{-0.019}\Big|_{\sigma_1}{}^{+0.001}_{-0.062}\Big|_{\mu}\nonumber\\
& {}^{+0.010}_{-0.004}\Big|_{M^2} {}^{+0.016}_{-0.017}\Big|_{s_0}{}^{+0.153}_{-0.079}\Big|_{\varphi_{\pm}(\omega)},   \\
 f_{B\to \pi}^{0}(0)=&0.122\times\bigg[ 
1\pm0.07\Big|_{S_0^\pi} {} \pm0.11\Big|_{\Lambda_q} \nonumber\\
& \quad \pm0.02\Big|_{\lambda_E^2/\lambda_H^2}
 {}^{+0.05}_{-0.06}\Big|_{{M^2}}  \pm0.05\Big|_{2\lambda_E^2+\lambda_H^2}\nonumber\\
&\quad ^{+0.06}_{-0.10}\Big|_{\mu_h } \pm0.04\Big|_{\mu} {}
^{+1.36}_{-0.56}\Big|_{\lambda_{B}}{}^{+0.25}_{-0.43}\Big|_{ \sigma_1, \sigma_2}\bigg]. 
\end{align}
It is evident that uncertainties from   heavy meson LCDAs,  namely the terms with the subscript $\lambda_B$, $\sigma_1$   and ${\varphi_{\pm}(\omega)}$  in the $B\to K^*$ form factor  and  the terms with the subscript $\lambda_B$ and  $\sigma_1, \sigma_2$    in the $B\to \pi$ form factor,   are dominant.  
Therefore, obtaining a thorough and reliable understanding of heavy meson LCDAs is essential for improving the precision of predictions in current research endeavors within the realm of heavy flavor physics.

However, to establish a reliable result on the full distribution of heavy meson LCDAs from the first-principle is extremely difficult due to various reasons. Firstly, these quantities are defined on the light cone, making them difficult to handle directly in first-principle calculations like lattice QCD. Secondly, the heavy meson LCDAs are formulated with the HQET field $h_v$, which also poses challenges for direct simulations on the lattice. Additionally, the simultaneous  presence of light-cone separation and the HQET effective field give rise to cusp divergences, which can be illustrated by a perturbative one-loop result for the relevant operators~\cite{Braun:2003wx}: 
\begin{align}
& O_v^{P, \mathrm{ren}}(t,\mu) = O_v^{P, {\rm bare}}(t) 
 \nonumber\\
&\;\;\;\; + \frac{\alpha_sC_F}{4\pi} \bigg\{\left(\frac{4}{\hat \epsilon^2} + \frac{4}{\hat \epsilon}\ln (it\mu)\right) O_v^{P, {\rm bare}}(t)  \nonumber\\
&\;\;\;\; - \frac{4}{\hat \epsilon}\int_0^1 du \frac{u}{1-u}[O_v^{P, {\rm bare}}(ut)-O_v^{P, {\rm bare}}(t)  ] \bigg\},
\end{align}
with  $d=4-\epsilon$ and the standard notation $2/\hat \epsilon=2/\epsilon -\gamma_E +\ln (4\pi)$.  The term ${4}\ln (it\mu) / {\hat \epsilon}$  arises from an expansion of ${4 (it\mu)^\epsilon}/{\hat \epsilon^2}$, which arises from the overlap of UV divergence and cusp divergence~\cite{Korchemskaya:1992je}. One can see that in the local limit $t \to 0$, the $\ln(it\mu)$ term also diverges.
Thus, the operator product expansion (OPE) for heavy meson LCDAs is ineffective, leading to ambiguities in defining the non-negative moments of heavy meson LCDAs. The conventional method for calculating these  moments also proves to be unsuccessful in this scenario, unlike for  light meson LCDAs.

In this work, we advance the approach proposed in Ref.~\cite{Han:2024min} to use  a sequential matching method to   circumvent all  three obstacles and determine heavy meson LCDAs from lattice QCD. To be explicit, one  first employs  the equal-time correlation functions, also named quasi distribution amplitudes (quasi-DAs), of a heavy meson with a large momentum component $P^z$ and the mass $m_H$.  By assigning a hierarchical ordering  $\pi/a \gg P^z \gg m_H \gg \Lambda_{\rm{QCD}}$,  dynamics associated with these scales can be separated by integrating out $P^z$ and $m_H$ in two steps.  Integrating out $P^z$, one can match the quasi-DAs to QCD LCDAs, which is done routinely in large momentum effective theory (LaMET)~\cite{Ji:2013dva,Ji:2014gla} (see Ref.~\cite{Ji:2020ect,Cichy:2018mum} for recent reviews).  After integrating out $m_H$, one can match the QCD LCDAs onto boosted HQET and obtain the required  LCDAs in HQET~ \cite{Ishaq:2019dst,Beneke:2023nmj}.  In this work, we will provide theoretical supports for this approach by demonstrating the rationale behind requiring this hierarchical ordering for the three involved scales, discussing the factorization at each step,   clarifying the underlying reason for obtaining HQET LCDAs in the final phase,  and addressing potential theoretical challenges.

Moreover  we will in this study make use of these theoretical advancements and  conduct a lattice QCD simulation of quasi-distribution amplitudes on a lattice ensemble with the lattice spacing $a = 0.05187$ fm to validate our proposed approach numerically. The gauge configurations employed are generated by the Chinese Lattice QCD (CLQCD) collaboration using $N_f=2 + 1$ flavor stout smeared clover fermions and Symanzik gauge action \cite{Hu:2023jet}. To enhance the stability of the pion mass at a given bare quark mass, a step of Stout link smearing is applied to the gauge field utilized by the clover action. 
With 549 gauge configurations, we perform 8784 measurements of the bare matrix elements of a heavy charmed $D$ meson with momenta $P^z=\{0,~6,~7,~8\}\times2\pi/La\simeq\{0,~2.99,~3.49,~3.98\}$ GeV. To improve the signal-to-noise ratio, we employ the grid source technique on the Coulomb gauge fixed configurations, involving a summation over all spatial sites along the $x$ and $y$ directions for each time slice. We  will employ three  fitting strategies to extract the bare matrix elements and eliminate contributions from excited states. For renormalization purposes, we utilize the state-of-the-art techniques and apply hybrid renormalization schemes at short and long distance separations. To mitigate long-distance perturbations, we perform an extrapolation in $\lambda= z\cdot P^z$ and assess the stability under variation of  various parameters.

After two-step matching, our results for HQET LCDAs  are found in agreement with existing phenomenological models. To facilitate their phenomenological application, we directly provide the tabulated  results for HQET LCDA  in the peak region. These results enable extrapolation to the low $\omega$ region using a model-independent parametrization. Subsequently, we determine the first inverse moment $\lambda_B$, along with the first and second inverse logarithmic moments $\sigma^{1,2}$ based on these results. Additionally, we fit the first inverse moment using recent models for HQET LCDAs, resulting in $\lambda_B \simeq 0.31-0.44$ GeV, which falls within the experimentally constrained range $\lambda_B > 0.24$ GeV established from the $B \to \gamma \ell\nu_\ell$ measurement \cite{Belle:2018jqd}. These predictions align with phenomenological determinations, demonstrating consistency.   Consequently, using the analytical results in Ref.~\cite{Gao:2019lta} we update the $B\to K^*$ form factors and show the dependence on the first inverse moment we have studied. 
The overall agreement of the pertinent results underlines the potential of the approach from Ref.~\cite{Han:2024min}  in providing first-principle predictions for heavy meson LCDAs, enhancing the prospects for phenomenological applications.

The remainder of this paper is structured as follows. In Sec.\ref{sec:theory}, we lay the foundations of the framework used, and explain how the proposed sequential effective theory method can be used to calculate the HQET LCDAs.  Sec. III provides details on the lattice QCD simulation, discussing the dispersion relation and fitting strategy. Sec. IV presents our calculations of quasi DAs, QCD  LCDAs, and HQET LCDAs. Sec. V delves into the phenomenological implications of our results. The concluding section provides a summary and a perspective on future enhancements.

\section{Framework for the Sequential effective theory} \label{sec:theory}

The presence of light-cone divergence is one of the reasons that prevent a direct computation of HQET LCDAs~\cite{Braun:2003wx}. In HQET, the heavy quark behaves akin to a Wilson line along the $v$ direction.  The coefficient for UV divergences between two Wilson lines is proportional  to the angle between these two Wilson lines which follows the definition~\cite{Korchemskaya:1992je}: 
\begin{eqnarray}
\cosh\theta = \frac{n_+\cdot v}{\sqrt{n_+^2}{\sqrt{v^2}}},
\end{eqnarray}
in which  $v^2=1$. Apparently for HQET LCDA, 
light-cone divergence, $n_+^2=0$, directly leads to  the cusp divergence. 

To address the divergence, two  directions can be considered in general. One option is to utilize an off-lightcone Wilson line, with some advantages discussed in Refs.~\cite{Kawamura:2018gqz, Wang:2019msf,Zhao:2020bsx,Xu:2022krn,Xu:2022guw,Hu:2023bba,Hu:2024ebp}. However, simulating heavy quark field on the lattice in this approach remains challenging, and successful applications have not yet been achieved.  The other option is to refrain from employing the HQET field $h_v$, and directly constructs the LCDAs with QCD  heavy quark field. In order to implement this approach successfully, it is essential to ensure that the infrared behavior of the two quantities involved are identical, with their differences being only at perturbative scale ${\cal O}(m_H)$~\cite{Han:2024min}. 
An expansion by region analysis of QCD LCDAs and HQET LCDAs  can be employed to confirm this fact~\cite{Beneke:2023nmj,Deng:2024dkd}. This necessitates boosting the heavy meson and employing a factorization to separate the hard-collinear  and soft-collinear degrees of freedoms. In this approach, lattice simulations must avoid using lightcone quantities which can be  achieved by using an equal-time correlator with a large momentum $P^z$ in LaMET. Following the factorization of hard and collinear degrees of freedom and taking into account the ultraviolet distinctions, one can convert  the equal-time correlator to QCD LCDAs as a low-energy effective theory.   Consequently, we succeed in addressing  the cusp divergence  together with the other two obstacles  by utilizing a quasi distribution amplitude that encompasses three scales with the hierarchical ordering:  $P^z \gg m_H \gg \Lambda_{\textrm{QCD}}$~\cite{Han:2024min}.

\begin{figure*}
  \centering
  \includegraphics[width=0.7\linewidth]{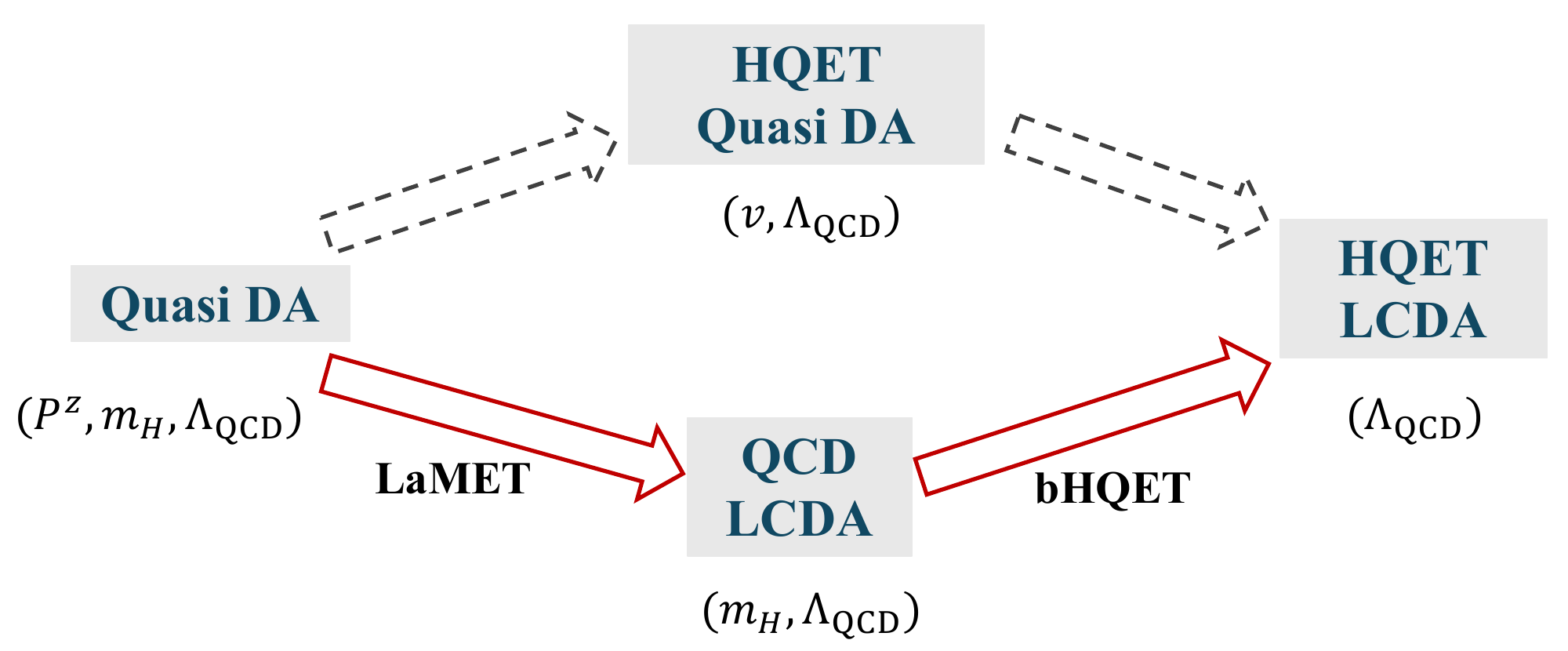}
  \caption{The procedure to calculate the HQET LCDAs. We adopt a two-step matching scheme which makes sequential use of effective field theories.   In this approach,  one starts with  the quasi DAs.  After factorizing the hard and collinear degrees of freedom, one can obtain the QCD LCDAs which contains the dynamics far below $P^z$. The collinear modes with offshellness $m_Q^2\sim m_H^2$ are integrated out and the QCD LCDAs can be matched onto the HQET LCDAs in  a boost frame. This procedure is depicted by the red arrows. Another alternative approach that makes use of quasi DAs within  HQET is also shown in this figure.  }
  \label{fig:two_step_scheme}
\end{figure*}

By progressively isolating the higher two scales $P^z$ and $m_H$ through the utilization of two effective theory factorization formulas via LaMET  and boosted HQET (bHQET) sequentially, we can deduce the LCDAs within HQET. A visual representation of this procedure is depicted in Fig.~\ref{fig:two_step_scheme}.  The alternative approach~\cite{Kawamura:2018gqz, Wang:2019msf,Zhao:2020bsx,Xu:2022krn,Xu:2022guw,Hu:2023bba,Hu:2024ebp} that makes use of HQET quasi DA is also incorporated in this figure.  This will be  beneficial when direct simulation of the HQET heavy quark field on the lattice is straightforward to implement.

It is necessary to stress  that using the quasi DAs have the potential to resolve all three obstacles in calculating HQET LCDAs, but the procedure must be properly designed. 
More explicitly,   in the large momentum limit, the quasi-DA involves three distinct scales: the large momentum $P^z$,  the heavy hadron mass $m_H$ which is at leading power equal to the heavy quark mass $m_Q$,  and  the hadronic scale $\Lambda_{\textrm{QCD}}$. We will point out that only for $P^z \gg m_H \gg \Lambda_{\textrm{QCD}}$, the first two energy scales fall within the perturbative regime and can be integrated out step by step. In the first step, the scale $P^z$ can be integrated out, facilitating the matching of the quasi-DA to the LCDA defined in QCD. This is guaranteed by the separation of hard modes with offshellness $\left(P^{z}\right)^2$  and collinear modes.  This procedure is consistent with the treatment of parton distribution functions (PDFs) and light meson LCDAs in LaMET~(for recent developments, please refer to the reviews~\cite{Ji:2020ect,Cichy:2018mum}), while a difference is that  collinear modes in the present case include both hard-collinear  and soft-collinear modes.   Once the QCD LCDA is derived, the second step involves integrating out the $m_H$ scale and matching the resulting quantity to the LCDA defined in HQET, more explicitly in boosted HQET. Hard-collinear modes with an off-shellness of $m_Q^2$  are absorbed by the jet function, while the low-energy contributions are accounted for in the HQET LCDAs~\cite{Ishaq:2019dst,Beneke:2023nmj}.   In the subsequent discussion, we will explain in detail who to carry out this two-step matching procedure.

\subsection{Step I: Integrating out $P^z$} \label{sec:theo_LaMET_matching}

\begin{figure}
  \centering
  \includegraphics[width=0.8\linewidth]{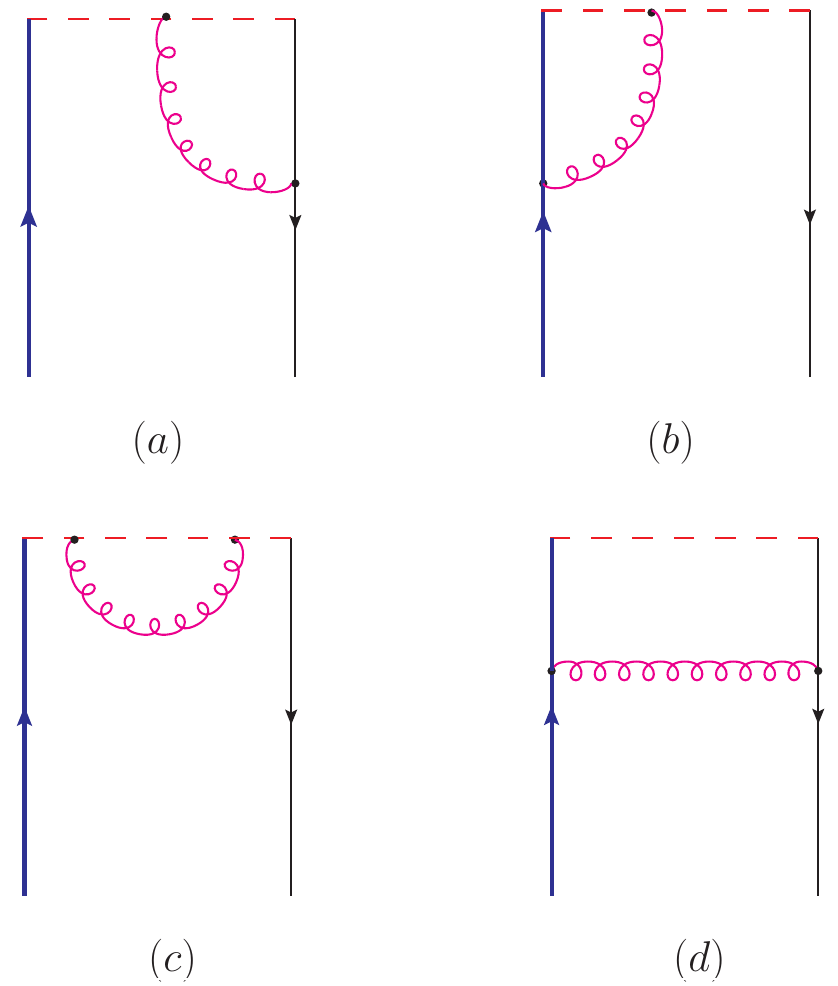}
  \caption{The one-loop Feynman diagrams quasi-DA in Eq.\,(\ref{eq:definitionofquasiDA}). The red dashed line represents the gauge link, while the blue and black line denote the heavy quark and light quark, respectively.}
  \label{fig:Feynman_diagram}
\end{figure}

LaMET calculation involves a separation of hard and collinear components~\cite{Ji:2013dva,Ji:2014gla}, and has a wide range of applications on the lattice, including calculations of quark distribution functions~\cite{Xiong:2013bka,Lin:2014zya,Alexandrou:2015rja,Chen:2016utp,Alexandrou:2016jqi,Alexandrou:2018pbm,Chen:2018xof,Lin:2018pvv,LatticeParton:2018gjr,Alexandrou:2018eet,Liu:2018hxv,Chen:2018fwa,Izubuchi:2018srq,Izubuchi:2019lyk,Shugert:2020tgq,Chai:2020nxw,Lin:2020ssv,Fan:2020nzz,Gao:2021hxl,Gao:2021dbh,Gao:2022iex,Su:2022fiu,LatticeParton:2022xsd,Gao:2022uhg,Chou:2022drv,Gao:2023lny,Gao:2023ktu,Chen:2024rgi,Holligan:2024umc,Holligan:2024wpv}, gluon distribution functions~\cite{Wang:2017qyg,Wang:2017eel,Fan:2018dxu,Wang:2019tgg,Good:2024iur}, generalized parton distributions~\cite{Chen:2019lcm,Alexandrou:2019dax,Lin:2020rxa,Alexandrou:2020zbe,Lin:2021brq,Scapellato:2022mai,Bhattacharya:2022aob,Bhattacharya:2023nmv,Bhattacharya:2023jsc,Lin:2023gxz,Holligan:2023jqh,Ding:2024hkz}, distribution amplitudes~\cite{Zhang:2017bzy,Chen:2017gck,Zhang:2020gaj,Hua:2020gnw, LatticeParton:2022zqc,Gao:2022vyh,Xu:2022guw,Holligan:2023rex,Deng:2023csv,Han:2023xbl,Han:2023hgy,Han:2024min,Han:2024ucv,Baker:2024zcd,Cloet:2024vbv,Han:2024cht,Deng:2024dkd,Chu:2024vkn}, transverse-momentum-dependent distributions~\cite{Ji:2014hxa,Shanahan:2019zcq,Shanahan:2020zxr,Zhang:2020dbb,LatticeParton:2020uhz,Ji:2021znw,LatticePartonLPC:2022eev,Liu:2022nnk,Zhang:2022xuw,Deng:2022gzi,Zhu:2022bja,LatticePartonCollaborationLPC:2022myp,Rodini:2022wic,Shu:2023cot,Chu:2023jia,delRio:2023pse,LatticePartonLPC:2023pdv,LatticeParton:2023xdl,Alexandrou:2023ucc,Avkhadiev:2023poz,Zhao:2023ptv,Avkhadiev:2024mgd,Bollweg:2024zet,Spanoudes:2024kpb,LPC:2024voq}, and double parton distribution functions~\cite{Zhang:2023wea,Jaarsma:2023woo}. Hard modes at the scale $\mu=P^z$ are captured in the short-distance coefficient, while the collinear sector, encompassing the low-energy degrees of freedom $(m_Q^2, \Lambda_{\rm QCD}^2)$, is accounted for within the LCDAs defined with QCD quark fields.

In LaMET,  the relevant quasi-DA is defined as
\begin{align}
\tilde{\phi}(x, P^z)=\int \frac{dz}{2\pi} e^{-ix P^z z}\tilde{M}^R(z,P^z). \label{eq:definitionofquasiDA}
\end{align}
In this context, $\tilde{M}^R(z,P^z)$ represents the renormalized nonlocal matrix element (ME) of a highly boosted heavy meson in coordinate space.  Here, we consider the renormalization of the bare ME $\tilde{M}^B(z, P^z; \Gamma)$ which is defined as:
\begin{align}
&\tilde{M}^B(z, P^z; \Gamma) =\frac{\left\langle 0\left|\bar{q}(z)\Gamma W_c(z,0) Q(0) \right|H(P^z)\right\rangle}{\left\langle 0\left|\bar{q}(0)\Gamma Q(0) \right|H(P^z)\right\rangle},
\label{eq:definitionofquasiDA2}
\end{align}
in the hybrid-ratio scheme \cite{Ji:2020brr}. The hybrid renormalization involves identifying and subsequently subtracting linear and logarithmic divergences present in bare  MEs. This renormalization procedure is carried out in two distinct regions \cite{LatticePartonLPC:2021gpi,Holligan:2023jqh,Baker:2024zcd}:
\begin{align}
\tilde{M}^{R}(z, P^z)= \left\{
\begin{array}{lr}
\frac{\tilde{M}^{B}(z, P^z;\gamma^z\gamma_5)}{\tilde{M}^{B}(z, P^z=0;\gamma^t\gamma_5)}, & |z|< z_s \\
e^{(\delta m+m_0)(z-z_s) }  \frac{\tilde{M}^{B}\left(z, P^z;\gamma^z\gamma_5\right)}{\tilde{M}^{B}(z_s, P^z=0;\gamma^t\gamma_5)}, & |z| \geq z_s
\end{array}
\right.
\label{eq:renormalizationform}
\end{align}
where $z_s$ represents a scale distinguishing the correlations at short and long distances, allowing for their separate renormalization. In principle, $z_s$ should be chosen within the perturbative region to ensure that the ratio scheme at short distances can be reliably converted to the perturbative $\overline{\mathrm{MS}}$ scheme.
$\delta m$ delineates   linear divergence within   bare matrix elements, stemming from self-energy of Wilson line and influencing the decay properties at large values of $z$. Additionally, $m_0$ encapsulates the regularization scheme-specific renormalon ambiguity, arising due to the non-convergence of the perturbation series used to compute $\delta m$ across all orders.

The logarithmic UV  divergence within the bare matrix elements, stemming from the  perturbative corrections to  the quark-Wilson-line vertex, can be eliminated by normalizing certain lattice MEs that exhibit the same divergence. For instance, MEs with the same operator but different external states can be leveraged for this purpose. 
In our study, we employ the zero-momentum MEs $\tilde{M}^B\left(z, P^z=0; \gamma^t\gamma_5\right)$ to renormalize the bare MEs with large momentum external state. It is important to note that the bare MEs in Eq.~(\ref{eq:definitionofquasiDA2}), featuring the Dirac structure $\gamma^{\mu}\gamma_5$, may suffer from both operator mixing effects \cite{Constantinou:2017sej,Chen:2017mzz,Chen:2017mie} due to chiral symmetry breaking  and power corrections from the terms proportional to $z^{\mu}$ \cite{Bhattacharya:2022aob,Cloet:2024vbv} in nonlocal MEs. Ref.\cite{Liu:2018tox} indicates that the operator mixing effects in pseudoscalar meson quasi-DAs can be avoided by choosing $\Gamma=\gamma^z\gamma_5$. However, the leading power contributions of zero-momentum MEs with $\gamma^z\gamma_5$, which are proportional to $P^z=0$, are equal to zero and thus unsuitable as renormalization factors. As a substitute, we utilize the zero momentum MEs with $\Gamma=\gamma^{t}\gamma_5$, which possess the same UV divergence as the bare large-momentum MEs with $\Gamma=\gamma^{z}\gamma_5$.

In the regime where the meson momentum $P^z$ substantially exceeds the   heavy meson mass, the renormalized quasi-DA $\tilde{\phi}(x, P^z)$ can be matched to the QCD LCDA via the following factorization formula
\begin{align}
\tilde{\phi}(x,P^z) =& \int_0^1 dy \, C \left(x, y, \frac{\mu}{P^z}\right) \phi(y, \mu) \nonumber\\
 & +\mathcal{O} \left(\frac{m_H^2}{\left(P^{z}\right)^2},\frac{\Lambda_{\mathrm{QCD}}^2}{(xP^z, \bar xP^z)^2}\right)
\label{eq:LaMETmatching}
\end{align}
with $\bar x= 1-x$. $\phi(y, \mu)$ denotes the QCD LCDA in $\overline{\textrm{MS}}$ scheme, which is defined as
\begin{align}
 & \phi(y, \mu) = \frac{1}{i f_{H}} \int_{-\infty}^{+\infty} \frac{d\tau}{2\pi} \, e^{i y P_H \tau n_+} \nn\\
  &\qquad \times \left\langle 0\left|\bar{q}(\tau n_+) {n}\!\!\!\slash_+\gamma_5 W_c(\tau n_+, 0) Q(0)\right| H(P_H)\right\rangle .  \label{defiofphiinQCD}
\end{align}
In the above formula $y$ represents the momentum fraction of the light quark in heavy meson $H$.  $|H(P_H)\rangle$ is the heavy meson state with mass $m_H$. $f_{H}$ signifies the decay constant in QCD. The $C \left(x, y, \frac{\mu}{P^z}\right)$ in Eq.(\ref{eq:LaMETmatching}) is the matching coefficient
\begin{align}
  C\left(x, y, \frac{\mu}{P^z}\right) 
  =~& \delta(x-y)+ C_{B}^{(1)}\left(x, y, \frac{\mu}{P^z}\right)   \nonumber\\
  &  - C_{CT}^{(1)}\left(x, y\right) +\mathcal{O}(\alpha_s^2),  \label{eq:matchingcoefficient}
\end{align}
where $C_B^{(1)}$ is the bare matching kernel, and $C_{CT}^{(1)}$ denotes the counter term in the hybrid-ratio scheme.

It is convenient to perform the calculation in Feynman gauge, and the dimensional regularization ($d\!=\!4\!-\!2\epsilon$) can be utilized to handle both ultraviolet and infrared divergences. The relevant  Feynman diagrams at one-loop are shown in Fig.~\ref{fig:Feynman_diagram}. At leading power, we find that 
the $C_{B}^{(1)}\left(x, y, \frac{\mu}{P^z}\right)$ is identical to the matching kernel for the light meson, which has been reported in Ref.~\cite{Liu:2018tox},
\begin{align}
 & C_{B}^{(1)}\left(x, y, \frac{\mu}{P^z}\right)\nonumber\\
  =& \frac{\alpha_s C_F}{2 \pi} \begin{cases}{\left[H_1(x, y)\right]_{+}} & x<0<y \\ {\left[H_2\left(x, y, P^z / \mu\right)\right]_{+}} & 0<x<y \\ {\left[H_2\left(1-x, 1-y, \frac{P^z}{ \mu}\right)\right]_{+}} & y<x<1 \\ {\left[H_1(1-x, 1-y)\right]_{+}} & y<1<x \end{cases} \,, 
 \end{align}
where
\begin{align}
  H_1(x, y) &=\frac{1+x-y}{y-x} \frac{1-x}{1-y} \ln \frac{y-x}{1-x} \nonumber\\
 &+\frac{1+y-x}{y-x} \frac{x}{y} \ln \frac{y-x}{-x} \,,\nn\\
 H_2\left(x, y, P^z / \mu\right) &=\frac{1+y-x}{y-x} \frac{x}{y} \ln \frac{4 x(y-x)\left(P^z\right)^2}{\mu^2} \nonumber\\
 & +\frac{1+x-y}{y-x}\left(\frac{1-x}{1-y} \ln \frac{y-x}{1-x}-\frac{x}{y}\right) \nn\,.
\end{align}
In the above formulas, the plus function is defined as
\begin{align}
  \left[ h(x,y) \right]_+ = h(x,y) - \delta(x-y)\int dz h(z,y).
\end{align}

In the hybrid scheme~\cite{Ji:2020brr}, a perturbatively controlled short-distance correction is introduced to eliminate the singularity at $z^2 \rightarrow 0$ to reconcile perturbation theory and lattice data. The Fourier transform of this correction into momentum space yields the counter-term $\tilde{\phi}_{CT}(x, x_0, P^z)$,
\begin{align}
&\tilde{\phi}_{CT}^{(1)}(x, P^z; x_0) \nonumber\\
&= \int \frac{d \lambda}{2\pi} e^{-i \lambda (x-x_0)}\frac{\alpha_s C_F}{2\pi} \frac{3}{2} \ln\left(\frac{\lambda^2}{z_s^2 P_z^2}\right) \theta\left(z_s-\left|\frac{\lambda}{P^z}\right|\right) \nonumber\\
&= - \frac{3\alpha_s C_F}{4\pi} \frac{2 \, \text{Si}[(x-x_0) z_s P^z] }{\pi (x-x_0)} \ ,
\end{align}
where $\text{Si}$ denotes the sine integral function. Accordingly, we can determine the counter-term $C_{CT}^{(1)}\left(x, y\right)$:
\begin{align}
C_{CT}^{(1)} = - \frac{3\alpha_s C_F}{4\pi} \left[\frac{2 \, \text{Si}[(x-y) z_s P^z] }{\pi (x-y)}\right]_{+} \ ,
\end{align}
where the plus function is introduced to preserve the proper normalization of the distribution. Notably, delta functions are excluded from the above derivation as they vanish within the overall plus function. It should be noted that the counter term  is derived  in the small spatial separation $z$ and higher power corrections are neglected. Inclusion of power corrections $m_Q \times z$ in the future are likely to improve the convergence of perturbation theory, and deserves a detailed analysis.  

\subsection{Step II: Integrating out $m_H$} \label{sec:HQETmatching}

It is important to note that in HQET, the momentum of a light quark inside a heavy meson at rest is typically soft.  Conversely, when the momentum of the light quark is large, with $\omega\sim m_H\gg \Lambda_{\mathrm{QCD}}$, the heavy quark will carry a relatively small momentum, referred to as the tail region. In this regime, the HQET LCDA is perturbatively calculable. Thereby it can be handled using QCD perturbation theory, and the one-loop result can be found in Ref.~\cite{Lee:2005gza}
\begin{align} \label{eq:tail_varphi_+}
 \varphi^+_{\mathrm{tail}}(\omega,\mu) = \frac{\alpha_s C_F}{\pi \omega} \bigg[ \bigg(\frac{1}{2}-\ln \frac{\omega}{\mu} \bigg)  + \frac{4\bar \Lambda}{3\omega} \left(2-\ln\frac{\omega}{\mu}\right)\bigg],
\end{align}
with $\bar \Lambda= m_H-m_Q$ characterizing the size of   power corrections.

When  the $\omega$ is small, this is the so-called peak region. In this regime,  the QCD LCDA can be matched  onto the HQET LCDAs using boosted HQET. Such a matching can be performed by employing heavy quark expansion with $m_H\gg \Lambda_{\mathrm{QCD}}$. It leads to a factorization theorem that the LCDA in QCD can be factorized into a multiplication of jet function $\mathcal{J}$ and the HQET LCDA  $\varphi^+(\omega,\mu)$. This relation was first derived for the inverse moment \cite{Pilipp:2007sb} and was later generated to LCDAs \cite{Ishaq:2019dst,Beneke:2023nmj} and the heavy-light QCD operators \cite{Zhao:2019elu}.

In this work  we  will adopt the result reported in Ref. \cite{Beneke:2023nmj}, in which the matching relation is written as
\begin{align}\label{eq:peak_varphi_+}
	\varphi^+_{\mathrm{peak}}(\omega, \mu)=\frac{1}{m_H} \frac{f_H}{\widetilde{f}_H} \frac{1}{\mathcal{J}_{\mathrm{peak}}}
	\phi(y, \mu; m_H)  
\end{align}
where the perturbative result for $\mathcal{J}_{\text {peak }}$ is given as: 
\begin{align}
	\mathcal{J}_{\text {peak }} =1&+\frac{\alpha_s C_F}{4\pi}\bigg( \frac{1}{2}\ln^2\frac{\mu^2}{m_H^2}+\frac{1}{2}\ln\frac{\mu^2}{m_H^2}+\frac{\pi^2}{12}+2\bigg) \nn\\
	&+\mathcal{O}\left(\alpha_s^2\right), 
\end{align}
with $\omega = y m_H$, $\bar y \equiv 1-y$. $f_H$ and $\widetilde{f}_H$ are the decay constants of  heavy meson in QCD and HQET, respectively. Their relation is given by
\begin{align}
	f_H=\tilde{f}_H(\mu)\left[1-\frac{\alpha_s C_F}{4 \pi}\left(\frac{3}{2} \ln \frac{\mu^2}{m_Q^2}+2\right)+\mathcal{O}\left(\alpha_s^2\right)\right].
\end{align}

The factorization formula in Eq.~(\ref{eq:peak_varphi_+}) is multiplicative. The scale $\mu$ here satisfies $\Lambda_{\rm QCD}\le \mu\le m_H$, and the renormalization group equation allows to evolve the HQET LCDAs to a low energy scale.

With the results in Eq.~\ref{eq:tail_varphi_+} and Eq.~(\ref{eq:peak_varphi_+}) for the tail and peak regions repsectively, we can merge them to obtain a full distribution: 
\begin{align}\label{eq:matching_between_QCD_HQET}
 \varphi^+(\omega, \mu)=\left\{\begin{array}{lr}
	\varphi^+_{\mathrm{peak}}(\omega, \mu),~~~~&\omega \sim \Lambda_{\mathrm{QCD}}\\\\
	\varphi^+_{\mathrm{tail}}(\omega, \mu).~~~~&\omega \sim m_H
	\end{array}\right.
\end{align}
The details of merging these two regions will be discussed in Sec.\ref{sec:DeterminHQETLCDA}.

\subsection{Theoretical  clarifications  }

In the preceding subsections, we have detailed the two-step matching method that involves the successive application of effective field theories. Beyond the theoretical methodology, there are several noteworthy observations to be highlighted.
\begin{itemize}
\item In heavy meson decays, the HQET LCDAs defined in the rest frame with $ (v^\mu=(1,0,0,0)) $
 are typically employed. However, in this study, the QCD LCDAs can be matched onto the LCDAs defined in the boosted HQET scenario~\cite{Fleming:2007qr,Fleming:2007xt}: 
\begin{align}
   \langle 0|\mathcal{O}_b^P(\omega) |H(p_H)\rangle =& -i \tilde f_H    \varphi^+(\omega;\mu), 
\end{align} 
The involved  bHQET operator is constructed as: 
\begin{align}
\mathcal{O}_b^P&(\omega)=\frac{1}{m_H}\int\frac{dt}{2\pi}e^{-it\omega n_+\cdot v}\nonumber\\
&\times  \sqrt{\frac{n_+\cdot v}{2}}\bar \xi_{sc}(tn_+) \slashed{n}_+\gamma_5W_c[tn_+,0]{h}_n(0), 
\end{align}
where the soft-collinear field $\xi_{sc}=\frac{\slashed{n}_- \slashed{n}_+}{4}q_{sc}(x)$ describes the light anti-quark in the heavy meson in the boosted frame. The bHQET field $h_n$  \cite{Dai:2021mxb} is given as: 
\begin{align}
\label{eq:HQETfield}
h_n(x)\equiv \sqrt{\frac{2}{n_{+}\cdot v}}e^{i m_Q v \cdot x}\frac{\slashed{n}_-\slashed{n}_+}{4}Q(x), 
\end{align}
which is considered as a hard-collinear field.  An  analysis based on expansion by region has demonstrated that the matching of QCD LCDAs to HQET LCDAs is the same in these two frames~\cite{Deng:2024dkd}. 

\item  It is important to note that HQET LCDAs are not dependent on   heavy quark mass. Therefore, in the methodology, one has the flexibility to choose $m_Q$ as the charm quark mass $m_c$, the bottom quark mass $m_b$, or any other suitable value, with the resulting HQET LCDAs remaining unchanged. Discrepancies observed in the obtained results can indicate the presence of power corrections. In the following simulation we will choose $m_Q=m_c$.

\item Given that HQET LCDAs are not influenced by the heavy quark mass, the mass dependence of QCD LCDAs can be deduced, leading to the derivation of an evolution equation for this mass dependence~\cite{Wang:2024wwa}. This process aligns with the concept of a renormalization group equation, which can enhance the power expansion in relation to $ (\Lambda_{\text{QCD}}/m_Q)$.   Furthermore, employing QCD LCDAs to derive HQET LCDAs aligns with the principle of LaMET, which illustrates that lightcone observables, traditionally defined in the infinite momentum limit, can be accessed using a finite yet large $P^z$.  
Likewise, exploring observables related to heavy quarks defined in HQET is achievable through utilizing quantities that involve a large yet finite heavy quark mass.

\item In   heavy quark limit  $(m_Q\to \infty)$, the interaction becomes uncorrelated with   heavy quark spin, establishing the presence of heavy quark spin symmetry. As a result, the identical HQET LCDAs can be defined for a heavy vector meson: 
\begin{align}\label{eq:HQETLCDA_spin_symmetry} 
    \langle 0| O_{v}^{||}(tn_+) |H^*(p_H,\eta)\rangle =& \tilde  f_H  m_H n_+ \cdot \eta  \nonumber\\
    \times \int_0^\infty d\omega e^{i \omega t n_+\cdot v}& \varphi^+(\omega;\mu),\nonumber\\
    \langle 0| O_{v}^{\perp\mu}(tn_+) |H^*(p_H,\eta)\rangle =& 
    \tilde  f_H  m_H n_+\cdot v \eta_{\perp}^{\mu}\nonumber\\
    \times \int_0^\infty d\omega e^{i \omega t n_+\cdot v}& \varphi^+(\omega;\mu),
\end{align} 
where $\eta$ is the polarization vector. The involved  operators are given as:  
\begin{align} 
O_{v}^{||}(tn_+) =& \, \bar q_s(tn_+) \slashed{n}_+W_c[tn_+,0]{h}_v(0), \nonumber\\
 O_{v}^{\perp\mu}(tn_+)=& \, \bar q_s(tn_+) \slashed{n}_+ \gamma^\mu_{\perp}W_c[tn_+,0] {h}_v(0). 
\end{align}
Building upon this symmetry, a recent study demonstrates that quasi distribution amplitudes for a vector meson can also be employed to extract the same HQET LCDAs~\cite{Deng:2024dkd}. Discrepancies in these extractions arise from power corrections attributed to the breaking of heavy quark spin symmetry.

\item  The factorization scheme in the present work  operates at the leading power level.  In the first step the factorization of hard and collinear modes, the power expansion is conducted  in terms of $m_H^2/\left(P^{z}\right)^2$, and $\Lambda_{\mathrm{QCD}}^2/(xP^{z})^2, \Lambda_{\mathrm{QCD}}^2/((1-x)P^{z})^2$.  Some power corrections, target mass corrections with $m_H^2/\left(P^{z}\right)^2$, have been estimated, and $\Lambda_{\mathrm{QCD}}^2/(xP^{z})^2$ corrections are estimated using a renormalon model~~\cite{Han:2024cht}. The findings suggest that these corrections typically fall below $20\%$, but  it is important to note that other power corrections, particularly those in the subsequent step involving terms of $\Lambda_{\mathrm{QCD}}/m_Q$, have not yet been explored.

\item The two-step matching method derives the HQET LCDAs  only  in the peak region. However, in order to integrate the perturbative result in the endpoint region, a strategy for their combination must be devised.
In bringing together these outcomes, it is crucial to identify a region where both predictions are anticipated to be accurate. A suitable selection for this integration is around $\omega \sim 1$GeV.

\item  In LaMET, the prediction for QCD LCDAs is uncertain in the region where $y\sim \Lambda_{\mathrm{QCD}}/P^z$. This translates to $\omega= ym_H\sim \Lambda_{\mathrm{QCD}}/P^z\times m_H$. For typical values of $P^z\sim (3-4)$ GeV and $m_H\sim (1.5-2)$
 GeV, this corresponds to $y\sim0.1$ and $\omega\sim 0.2$ GeV, which falls below the peak region of HQET LCDAs. 
To match such high momenta, lattice ensembles with a lattice spacing of $a\lesssim 0.05$ fm are required and will be used in our lattice simulation.

\end{itemize}


\section{Lattice simulation}

\subsection{Lattice setup}

The numerical simulation in this work is performed on the gauge configuration generated by China Lattice
QCD (CLQCD) collaboration with $N_f=2 + 1$ flavor stout smeared clover fermions and Symanzik gauge action \cite{Hu:2023jet}.  One step of Stout link smearing is applied to the gauge field used by the clover action to improve the stability of the pion mass for given bare quark mass. 
In this work, we adopt a single ensemble (named ``H48P32'' in Ref.\cite{Hu:2023jet}) at the finest lattice spacing $a=0.05187$fm and volume $L^3\times T=48^3\times144$ with sea quark masses corresponding to $m_{\pi}=317.2$MeV and $m_K=536.1$MeV.
Furthermore, we determine the charm quark mass by requiring the corresponding $J/\psi$ mass to have its physical value $m_{J/\psi}=3.96900(6)$GeV within $0.3\%$ accuracy, hence the heavy $D$ meson mass on this ensemble is calibrated as $m_D=1.917(5)$GeV. These configurations    have been successfully applied in studies involving hadron spectrum~\cite{Liu:2022gxf,Xing:2022ijm,Yan:2024yuq}, decay and mixing of charmed hadron~\cite{Zhang:2021oja,Liu:2023feb,Liu:2023pwr,Meng:2024nyo,Du:2024wtr,Chen:2024rgi} and other interesting phenomena~\cite{Zhao:2022ooq,Meng:2023nxf}.

From a total of 549 gauge configurations, we use 8784 measurements of the bare MEs of heavy $D$ meson with boosted momentum $P^z=\{0,~6,~7,~8\}\times2\pi/(La)\simeq\{0,~2.99,~3.49,~3.98\}$GeV respectively. 
To improve the signals, we adopt the plane source on the Coulomb gauge fixed configurations, which involves summing over all spatial sites along the $x$ and $y$ directions for each time slice. The advantage of using this particular source lies in its ability to produce good correlation signals through the summation of spatial wave functions across all sites in the $x$ and $y$ directions. Simultaneously, the fixed positional coordinates along the $z$ axis guarantee the ability to obtain any momentum values following a Fourier transformation.

We use the grid-source-to-point-sink propagators 
\begin{align}
	G_f(x^3, t, t_0) \equiv \sum_{\vec{y}}\sum_{\vec{x}} D^{-1}_f(t,\vec{y};t_0,\vec{x}) e^{i\vec{p}\cdot(\vec{x}-\vec{y})},
\end{align}
where $D^{-1}_f(t,\vec{y};t_0,\vec{x})$ denotes the $f$-flavored quark propagator from $(t_0,\vec{x})$ to $(t, \vec{y})$. The summation of $\vec{x}=(x^1, x^2)$ corresponds to the summing over all spatial sites along  $x$ and $y$ directions of the grid source, and the residual $x^3$ will contribute to the momentum of the hadronic external state in the nonlocal two point function (2pt),
\begin{align}
	C_2(z,P^z,t; \Gamma)=&\sum_{x^3}e^{iP^zx^3}  \left\langle G_q(x^3+z,t,0)\Gamma  \right.\nn\\
	& \times \left. W_c(x^3+z,x^3)\gamma_5G_Q^{\dagger}(x^3,t,0)\right\rangle , \label{eq:nonlocal2pt}
\end{align}
where $z$ denotes the spatial separation between the light and heavy quarks within the nonlocal quark operator, and $W_c(x^3+z,x^3)$ is the spatial Wilson line along the $z$ direction that ensures gauge invariance. 
As mentioned above, we opt for $\Gamma=\gamma^z\gamma_5$ to establish the correlations corresponding to large momentum  MEs, and for the zero momentum MEs, we choose $\Gamma=\gamma^t\gamma_5$.

\subsection{Extraction of bare matrix elements}

By incorporating single-particle intermediate state, the nonlocal  two point correlation  function in Eq.~(\ref{eq:nonlocal2pt}) can be parameterized as:
\begin{align}
	C_2(z, P^z, t ; \Gamma)&=\sum_n \frac{1}{2E_n}e^{-E_nt} \langle n(P^z)|\bar{Q}(0)\gamma_5q(0)|0\rangle \nn\\
	&\times\left\langle0|\bar{q}(z)\Gamma W_c(z, 0)Q(0)|n(P^z)\right\rangle.
\end{align}
The local ME is related to the decay constant of   heavy meson in QCD:
\begin{align}
	\langle n(P^{\mu})|\bar{Q}\gamma_5q|0\rangle = if_{H^{(*)}}P^{\mu}.
\end{align}
After taking a ratio of nonlocal and local 2pt, we can extract the ground state MEs $\tilde{M}^B(z, P^z; \Gamma)$ from the following form:
\begin{widetext}
\begin{align}
	\frac{C_2(z, P^z, t {; \Gamma})}{C_2(0, P^z, t; \Gamma)} &= \frac{\left\langle 0\left|\bar{q}(z)\Gamma W_c(z,0) Q(0) \right|H(P^z)\right\rangle}{\left\langle 0\left|\bar{q}(0)\Gamma Q(0) \right|H(P^z)\right\rangle} + e^{-(E_1-E_0)t} \nn\\
	&  \times \frac{\left\langle0|\bar{q}(z)\Gamma W_cQ(0)|H_1^*(P^z)\right\rangle\left\langle 0\left|\bar{q}(0)\Gamma Q(0) \right|H(P^z)\right\rangle - \left\langle0|\bar{q}(z)\Gamma W_cQ(0)|H(P^z)\right\rangle\left\langle 0\left|\bar{q}(0)\Gamma Q(0) \right|H_1^*(P^z)\right\rangle}{\left|\left\langle 0\left|\bar{q}(0)\Gamma Q(0) \right|H(P^z)\right\rangle\right|^2} \nn\\
	& + e^{-(E_2-E_0)t} \cdots + \cdots \nn\\
	& \equiv \tilde{M}^B(z, P^z; \Gamma) \left[1 + \sum_{n=1}^{N_{\mathrm{stat}}-1} A_ne^{-(E_n-E_0)t} \right]_{N_{\mathrm{stat}}\to\infty}, \label{eq:paramterizationofR}
\end{align}
\end{widetext}
in which $H_n^*$ denotes the excited states of the pseudoscalar heavy meson $H$ with energy level $E_n$. One can extract $\tilde{M}^B(z, P^z; \Gamma)$ from a multi-state fit of ${C_2(z, P^z, t;\Gamma)}/{C_2(0, P^z, t;\Gamma)}$ according to the parametrization given by the last equation in Eq.(\ref{eq:paramterizationofR}), in which $A_n$ denotes the contribution from excited states and can be determined from fitting the $t$-dependence of the ratios.

\begin{figure}[http]
\centering
\includegraphics[width=\linewidth]{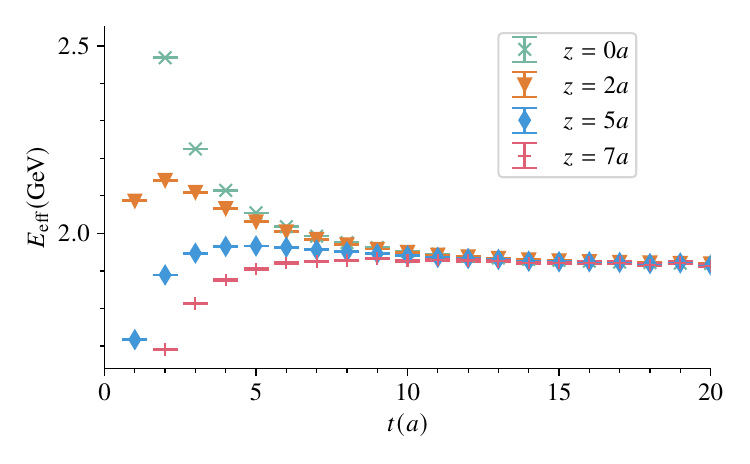}
\caption{Effective energies of nonlocal 2pt $C_2(z,P^z=0,t; \gamma^t\gamma_5)$ at different $z$, which is defined as $E_{\mathrm{eff}}=\ln C_2(z,P^z,t; \Gamma)/C_2(z,P^z,t+1; \Gamma)$ and will converge to $E_0$ at sufficiently large $t$. All the data tend to converge on the same plateau at large $t$, suggesting that they share the same ground-state energy. }
\label{fig:efftiveenergies}
\end{figure}

 \begin{figure}[htbp]  
	\centering
	\includegraphics[width=0.9\linewidth]{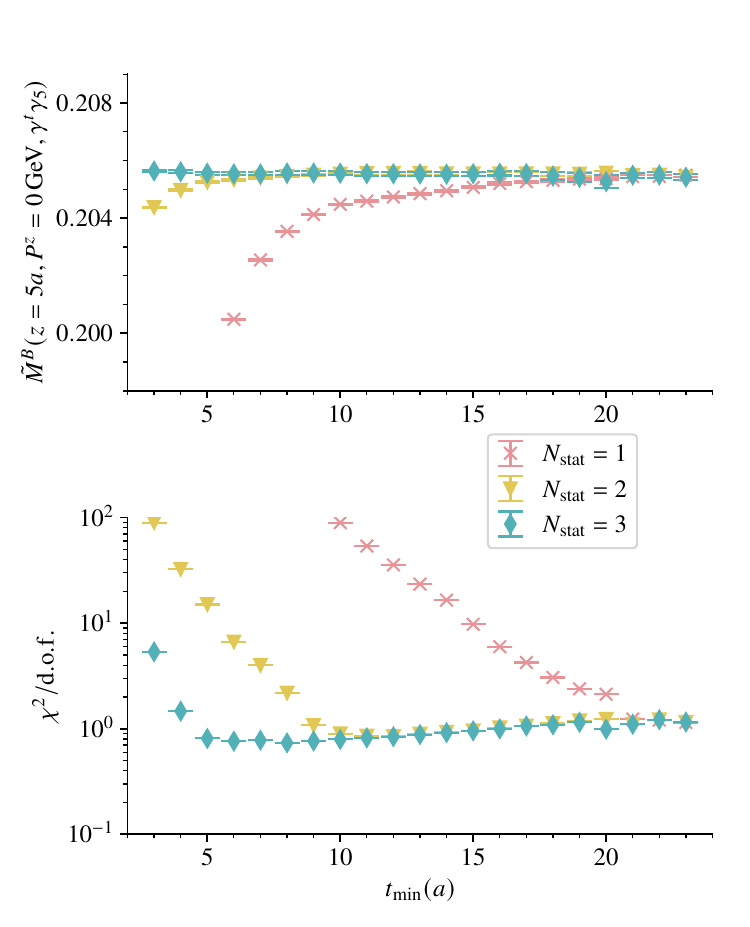}
	\caption{Upper panel: The fitting results of the ground-state ME $\tilde{M}^B(z,P^z=0;\gamma^t\gamma_5)$ from one-state, two-state, and three-state fit with a fit range $t\in[t_{\mathrm{min}}, 31a]$. Lower panel: The $\chi^2/$d.o.f. values of the fits.}
	\label{fig:compareNstatefits}
\end{figure}

We will conduct the fitting using one-state ($N_{\mathrm{stat}}=1$), two-state ($N_{\mathrm{stat}}=2$), and three-state ($N_{\mathrm{stat}}=3$) ansatz. 
First we exam the behavior of effective energies of the nonlocal 2pt $C_2(z,P^z=0,t; \gamma^t\gamma_5)$ at different $z$. As shown in Fig.~\ref{fig:efftiveenergies}, it is observed that all curves of $E_{\mathrm{eff}}(z)$  with the same external momentum $P$ exhibit a common plateau at large $t$.
To ensure the success of the one-state fits, it is necessary to utilize the data from sufficiently large $t$ values (e.g., $t\geq t_{\mathrm{min}}\simeq13a$ for $z=2a$ case, or $t_{\mathrm{min}}\simeq6a$ for $z=5a$ case in Fig.~\ref{fig:efftiveenergies}) to capture the plateau behavior. However, achieving this scenario is often challenging due to the signal-to-noise ratio in the nonlocal correlation functions, which tends to increase more rapidly for large momenta or spatial separations. This suggests that for the highly boosted nonlocal 2-point functions under consideration, the potential plateau behavior at large $t$ might be obscured by rapidly growing errors. As higher excited states possess larger energies, their contributions diminish quickly with the increase of $t$. Consequently, we can introduce the two-state, three-state (or even higher states) fits step by step by gradually reducing the value of $t_{\mathrm{min}}$ for each fit.

In the upper panel of Fig. \ref{fig:compareNstatefits}, we compare the results of the ground-state MEs $\tilde{M}^B(z,P^z=0;\gamma^t\gamma_5)$ obtained from one-state, two-state, and three-state fits with a fit range of $t\in[t_{\mathrm{min}}, 31a]$. It is evident that the results from all three fitting strategies stabilize and converge with each other when $t_{\mathrm{min}}>20a$. The lower panel displays the $\chi^2$/d.o.f. values, further confirming that in this region:  the $\chi^2$/d.o.f. values from the three fitting strategies are all slightly less than 1, indicating the reliability of the fits. When $t_{\mathrm{min}}$ is decreased from $20a$ to $10a$, deviations in the results from the one-state fits are observed, accompanied by an exponential increase in $\chi^2$ /d.o.f.. This behavior suggests that the contribution from the first excited state becomes significant, rendering a single-state parametrization inadequate to describe the data. Continuing to decrease $t_{\mathrm{min}}$, the effectiveness of the two-state fit diminishes as well, indicating the need to include more excited states to capture the more pronounced changes in the data.

\begin{figure}[htbp]
	\centering
	\subfigure[$P^z=2.99$GeV, $z=9a$]{\includegraphics[width=0.9\linewidth]{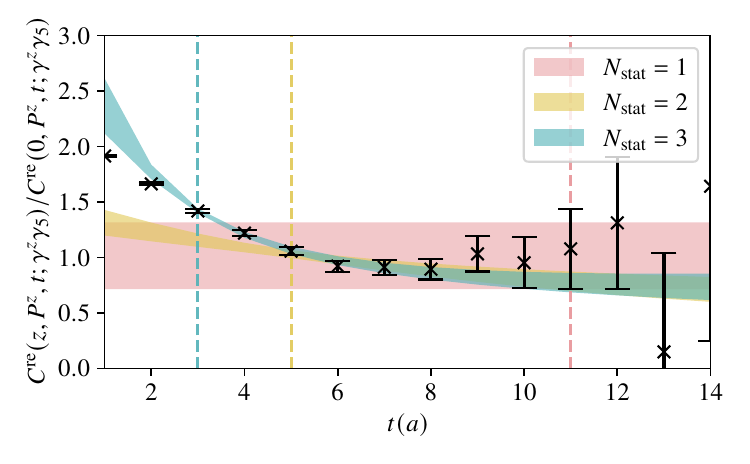}}
	\subfigure[$P^z=3.49$GeV, $z=5a$]{\includegraphics[width=0.9\linewidth]{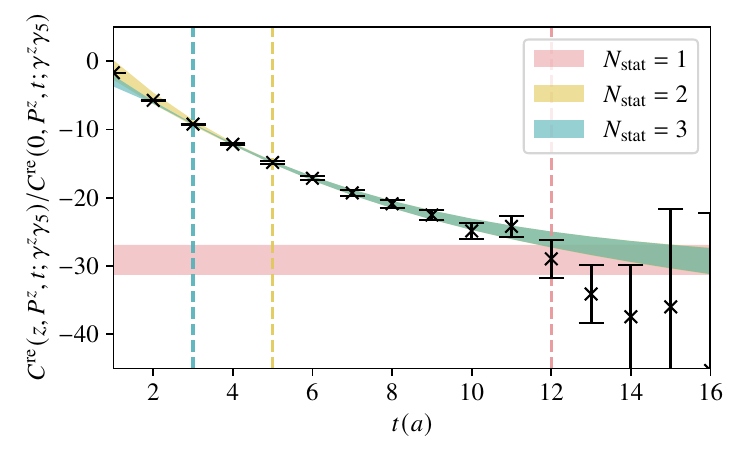}}
	\subfigure[$P^z=3.98$GeV, $z=10a$]{\includegraphics[width=0.9\linewidth]{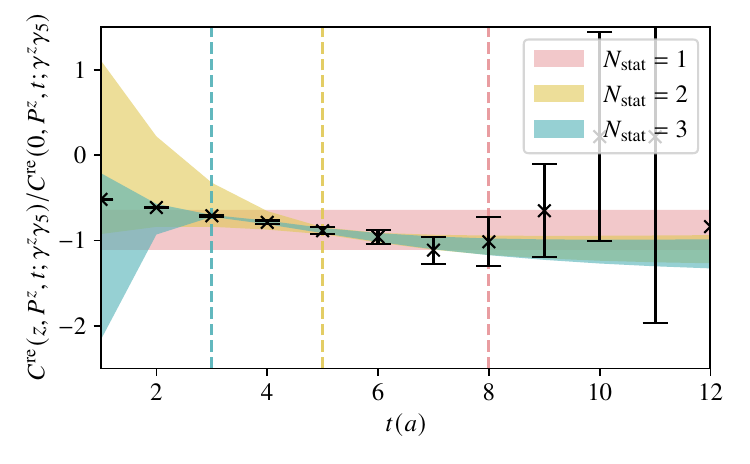}}
	\caption{Comparison of the original lattice data $C^{\mathrm{re}}(z,P^z,t;\gamma^z\gamma_5)/C^{\mathrm{re}}(0,P^z,t;\gamma^z\gamma_5)$ with large $P^z$ and $z$, along with the fit results of $\tilde{M}^B(z,P^z;\gamma^z\gamma_5)$ obtained using three fitting strategies. We choose the combinations for $(P^z,z)=(2.99\mathrm{GeV},9a),(3.49\mathrm{GeV},5a)$ and $(3.98\mathrm{GeV},10a)$ as examples to demonstrate the consistency of the results obtained from three fitting strategies. The dashed lines of varying colors represent the different $t_{\mathrm{min}}$  values utilized in each fit.}
	\label{fig:NstatefitsLargeMom}
\end{figure}

In practice, it is important to strike a balance between the number of excited states included and the fit range utilized in the fitting process. A conservative approach involves first performing the ground state fit over a sufficiently large range of $t$, after excluding  excited-state contaminations. However, this may not always be feasible, especially in situations where the signal-to-noise ratio increases rapidly. In such cases, it becomes necessary to judiciously reduce the value of $t_{\mathrm{min}}$  and incorporate the contributions from higher states to accurately describe the data.

In Fig.~\ref{fig:NstatefitsLargeMom}, we present a comparison of the original lattice data with large momentum $P^z$  and spatial separation $z$, along with the fit results obtained using three different strategies. The fits are carried out based on the parametrization form in Eq. (\ref{eq:paramterizationofR}). 
At large enough $t$, the data will converge to a plateau, corresponding to the ground state MEs. 
The various colored vertical dashed lines in the plots indicate the values of $t_{\mathrm{min}}$ utilized in the different fit strategies. It is evident that the signal-to-noise ratio of the original data experiences rapid growth and becomes unstable at large $t$. In such scenarios, there may be limited data available to effectively constrain the parameters when employing a one-state fit. By incorporating more excited states, we can utilize data from smaller $t$ values to better constrain the fit parameters. It is noticeable that the $t_{\mathrm{min}}$  values used for two-state and three-state fits gradually decrease, while the final fitting results from all three strategies remain consistent.

\begin{table}[htbp]
\caption{List of the fitting strategies and $t$ range utilized to extract all the large momentum MEs $\tilde{M}^B(z, P^z; \gamma^z\gamma_5)$. }
\renewcommand{\arraystretch}{2.0}
\setlength{\tabcolsep}{2mm}
\begin{tabular}{c c c c }
\hline\hline
   $z$ & $P^z=2.99$GeV & $P^z=3.49$GeV & $P^z=3.98$ GeV \\
	\hline 0 & 1-state, [11,16] & 2-state, [6,15] & 2-state, [6,13] \\
		1 & 1-state, [11,16] & 2-state, [6,15] & 2-state, [6,13] \\
		2 & 1-state, [11,16] & 2-state, [6,15] & 2-state, [6,13] \\
		3 & 1-state, [11,16] & 2-state, [5,15] & 2-state, [5,13] \\
		4 & 1-state, [11,16] & 2-state, [5,14] & 2-state, [5,13] \\
		5 & 1-state, [11,14] & 2-state, [5,13] & 2-state, [5,13] \\
		6 & 1-state, [11,14] & 2-state, [5,13] & 2-state, [5,13] \\
		7 & 1-state, [11,14] & 2-state, [5,12] & 2-state, [5,12] \\
		8 & 1-state, [11,14] & 2-state, [5,10] & 1-state, [8,12] \\
		9 & 1-state, [11,14] & 2-state, [5,9] & 1-state, [8,11] \\
		10 & 1-state, [11,14] & 2-state, [5,9] & 1-state, [8,11] \\
		11 & 1-state, [11,14] & 1-state, [8,13] & 1-state, [8,11] \\
		12 & 1-state, [10,14] & 1-state, [8,11] & 1-state, [8,11] \\
		13 & 1-state, [10,14] & 1-state, [8,11] & \\
		14 & 1-state, [10,14] & 1-state, [7,11] & \\
		15 & 1-state, [10,14] & & \\
		16 & 1-state, [10,14] & & \\
		\hline\hline
\end{tabular} 
\label{tab:FitStrategiesAndtRange}
\end{table}

\begin{figure}[htbp]
	\centering
	\includegraphics[width=1\linewidth]{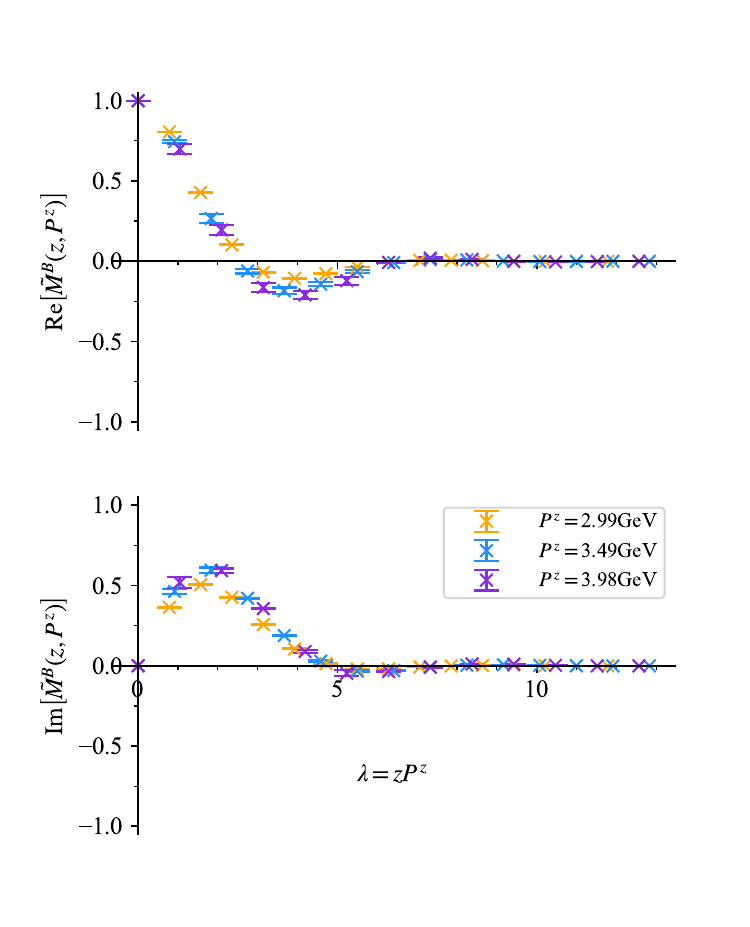}
	\caption{Bare MEs of quasi DA $\tilde{M}^B(z, P^z; \gamma^z\gamma_5)$ with boosted momenta $P^z=\{2.99,~3.49,~3.98\}$GeV. Both real and imaginary parts are shown as function of $\lambda=zP^z$.}
	\label{fig:bare_LaMME}
\end{figure}

In summary, the fitting strategies and $t$ ranges utilized to extract all $\tilde{M}^B(z, P^z; \gamma^z\gamma_5)$ are summarized in Tab. \ref{tab:FitStrategiesAndtRange}. For each fit, we initially assess the consistency of results obtained from different strategies, and subsequently determine the final fitting approach based on an evaluation of the fit quality. During the selection process, preference is given to the more conservative one-state fit. Only under conditions where the $\chi^2$ /d.o.f. of the fit exceeds 1, or when there are fewer than 4 effective data points available for fitting (to prevent overfitting), we resort to the two-state fit. The three-state fits are solely employed to validate the reliability of the results obtained from the initial two strategies and will  not be used to derive the final outcomes.

From the aforementioned deliberations, we ultimately derive the results  for the bare MEs of quasi DA $\tilde{M}^B(z, P^z; \gamma^z\gamma_5)$ with the boosted momenta $P^z=\{2.99,~3.49,~3.98\}$GeV, illustrated in Fig.~\ref{fig:bare_LaMME}.
Both real   and imaginary parts are shown as functions of $\lambda= z  P^z$  in this figure. 

\subsection{Dispersion relation}

\begin{figure}[htbp]
	\centering
	\includegraphics[width=0.9\linewidth]{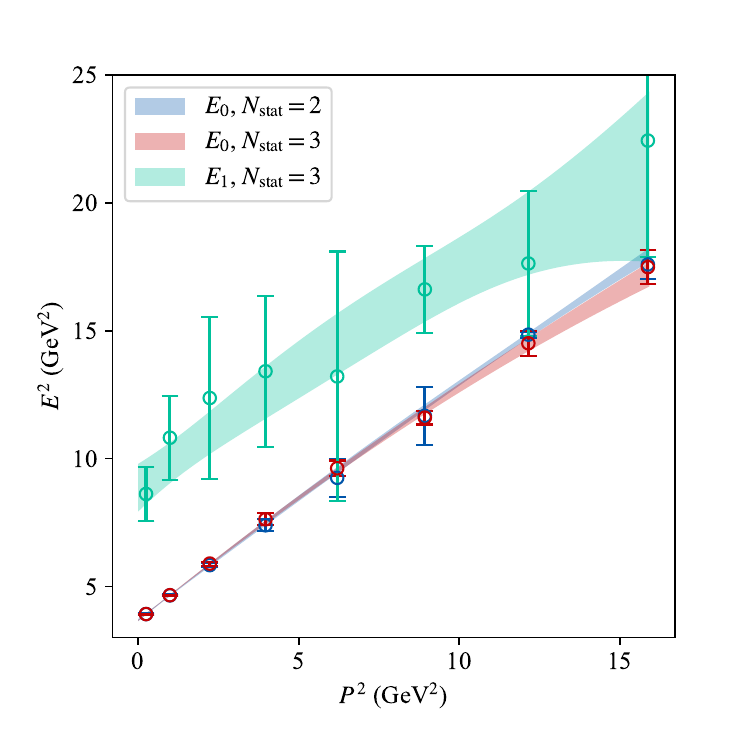}
	\caption{The dispersion relation of heavy $D$ meson with boosted momenta up to 3.98GeV. Two-state and three-state fits strategies are used to extract the ground-state (and first excited state from the latter one) energies. All the fit bands are consistent with the data points well.}
	\label{fig:disp_relation}
\end{figure}

Before delving into further discussions, it is essential to verify the dispersion relation for the heavy meson. The effective energies  with different momenta can be deduced from $R(z=0,P^z,t)$ with $\Gamma=\gamma^t\gamma_5$.  By applying the fitting methodologies outlined in the preceding subsection, we can acquire the outcomes for the ground state (from both two- and three-state fits) and the first excited state (from the three-state fit), as depicted in Fig. \ref{fig:disp_relation}.

For illustration, we adopt the following parametrization form to investigate the dispersion relation
\begin{align}
	E(P^z)=\sqrt{m^2+c_0\left(P^z\right)^2+c_1\left(P^z\right)^4a^2},
\end{align}
where   a quadratic term of the lattice spacing $a$ is introduced to account for discretization errors. The fit results are shown as the bands in Fig.\ref{fig:disp_relation}.  For the ground state energy $E_0$  obtained from the two-state fit, we find that $m=1.917(5)$GeV and $c_0=0.977(23)$, $c_1=-0.072 (25)$; for the ground state energy $E_0$ from three-state fit, we have $m=1.915(6)$GeV and $c_0=1.007(30)$, $c_1=-0.124(45)$. We have also conducted a fit for the dispersion relation of the first excited state, resulting in $E_1=2.98(16)$GeV, $c_0=1.00(43)$, $c_1=-0.21(49)$. 
These findings indicate that the results are fairly  consistent with the relativistic dispersion relation up to possible $P_z^2a^2$ discretization error. Therefore, for a moving  heavy meson  with momenta up to 4 GeV, the discretization effects we utilized remain controllable on this lattice ensemble.


\section{Numerical results for LCDAs}

\subsection{Renormalization in the hybrid scheme} \label{sec:RenormInRatioScheme}

Based on the renormalization formula in Eq.(\ref{eq:renormalizationform}), we will proceed with the renormalization of the bare MEs $\tilde{M}^B(z,P^z; \Gamma)$ in the hybrid-ratio scheme. 
As discussed in Sec.\ref{sec:theo_LaMET_matching}, an important concept in hybrid renormalization involves identifying and addressing the linear and logarithmic divergences in the bare MEs, followed by subtracting them in distinct regions.

In the long-distance region where $|z|\geq z_s$, which is defined in Eq.(\ref{eq:renormalizationform}), the renormalization factors will encompass additional non-perturbative effects arising from the conversion of lattice results obtained at a finite spacing $a$ to a continuum scheme. 
Their explicit origins are given as follows.  
\begin{itemize}
	\item $\delta m$ characterizes the linear divergence, which comes from the self-energy of the Wilson line in the bare ME, and governs the decay behavior at large-$z$. It can be extracted from fitting the long-range correlations of the zero momentum ME.
	\item $m_0$ has a complicated origin, including the regularization scheme dependent renormalon effect, pole mass, finite $P^z$ effects and fitting effect \cite{Ji:2020brr}. It can be determined from matching the lattice data of zero momentum MEs at small-$z$ range to the perturbative one.
\end{itemize}

\begin{figure}[htbp]
\centering
\includegraphics[scale=0.7]{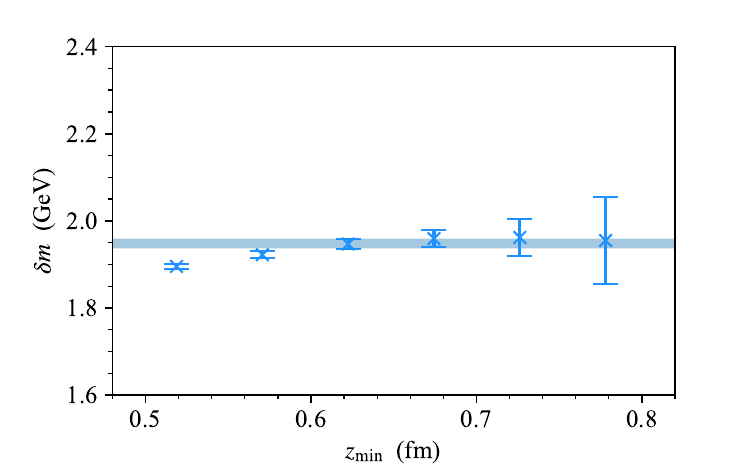}
\caption{Results of $\delta m$ by fitting the zero momentum MEs in the fit range $z\in[z_{\mathrm{min}}, z_{\mathrm{min}}+4a]$. It can be observed that the results of $\delta m$ converge after $z_{\mathrm{min}}>12a\simeq0.62$fm, then we obtain its value from a constant fit, as shown by the horizontal band.}
\label{fig:Ana_deltam_from_different_zmin}
\end{figure}

\begin{figure}[htbp]
\centering
\includegraphics[scale=0.7]{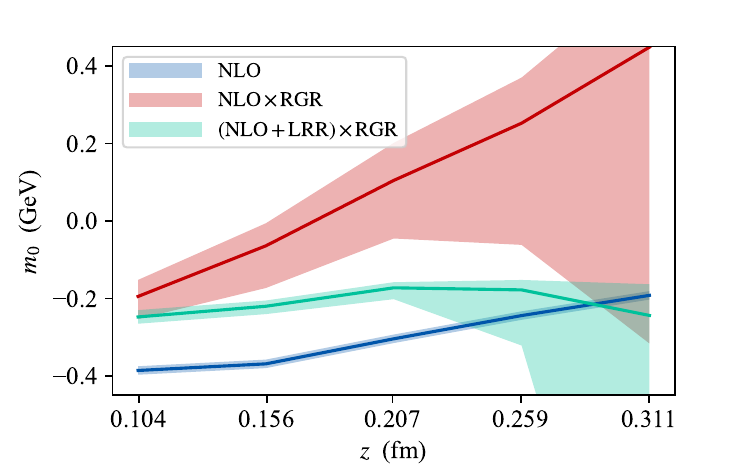}
\includegraphics[scale=0.7]{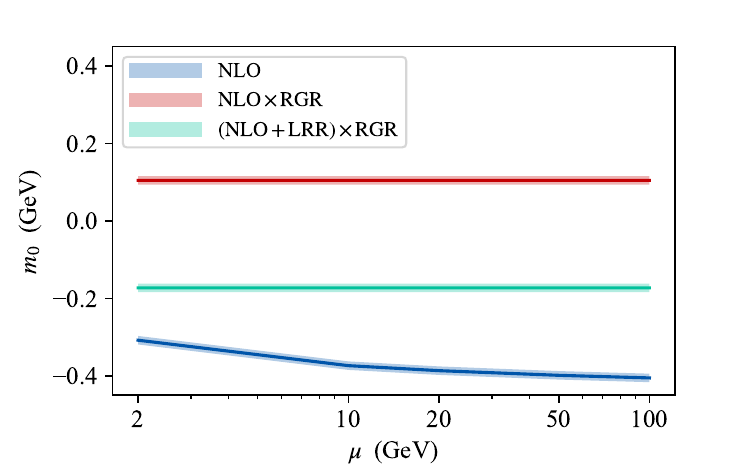}
\caption{Upper panel: Results of $m_0$ from Eq.(\ref{eq:fitformulaofm0}) with the perturbative Wilson coefficient schemes: ``NLO'', ``NLO$\times$RGR'' and ``(NLO+LRR)$\times$RGR''. The fitting is carried out in the range $[z-a,z+a]$ for each $z$.  Both statistical and systematic errors are included, where the latter one comes from the scale variation of $\mu_0$.  Lower panel: The scale dependence of $m_0$ from different schemes. Only statistical error is included in each bands. }
\label{fig:Ana_m0_dependence}
\end{figure}

The  $\delta m$ can be determined by fitting the MEs according to the exponential decay behavior $Ae^{-\delta mz}$ in the large-$z$ range~\cite{Ji:2020brr}. In principle, the parameter $\delta m$ should depend on the lattice spacing $a$ and the regularization scheme. However, with only one lattice spacing available at this stage, $\delta m$ is treated as a constant~\cite{Holligan:2023jqh}. Since $\delta m$ is independent of the momentum of the external state, it can be extracted by fitting the ME at zero momentum. In practice, we select the fitting range $z\in[z_{\mathrm{min}}, z_{\mathrm{min}}+4a]$ with $z_{\mathrm{min}}$ varying from $10a$ to $15a$ to assess the stability of the fits. The specific results obtained from different $z$-ranges are depicted as data points in Fig.~\ref{fig:Ana_deltam_from_different_zmin}. It is observed that the values of $\delta m$ obtained from different intervals in the large-$z$ region are consistently around 1.9 GeV, indicating the universality of the extracted $\delta m$.   By performing a constant fit of the $\delta m$ values from different $z_{\mathrm{min}}$ values, the exact value of $\delta m$ is determined to be $\delta m=1.948(11)$ GeV, as shown by the band in Fig.\ref{fig:Ana_deltam_from_different_zmin}.

After removing the linear divergence, there still remains an $a$-independent term $m_0$ collecting the renormalon ambiguity and other effects \cite{Ji:2020brr}. At small-$z$ region, where the perturbation theory works well, one can extract $m_0$ by matching the renormalized zero momentum ME to the continuum perturbative $\overline{\mathrm{MS}}$ result of the Wilson coefficient $C_0(z,\mu)$ \cite{Baker:2024zcd}
\begin{align}
 \tilde{M}^B(z, 0; a) &e^{\left(\delta m(a)+m_0\right) z}= \nonumber\\
 &\qquad C_0\left(z, \mu_0\right) e^{-\mathcal{I}\left(\mu_0\right)} e^{\mathcal{I}^{\mathrm{lat}}\left(a^{-1}\right)}. \label{eq:matchingRenormalization}
\end{align}
Here $\mathcal{I}(\mu)=\int_{\alpha_s(\mu_0)}^{\alpha_s(\mu)} d \alpha \frac{\gamma(\alpha)}{\beta(\alpha)}$ denotes the renormalization group resummation (RGR) improvement that cancels the renormalization scheme dependence of the results between lattice scale $\mu_0=2e^{-\gamma_E}z^{-1}$ and perturbative $\overline{\mathrm{MS}}$ scale $\mu$. The result of Wilson coefficient up to the next-to-leading order (NLO) in $\alpha_s$ reads
\begin{align}
	C_0^{\mathrm{NLO}}(z, \mu)=1+\frac{\alpha_sC_F}{2\pi}\left[\frac{3}{2}\log\left(\frac{1}{4}e^{2\gamma_E}z^2\mu^2 \right) + A \right], \label{eq:C0NLO}
\end{align}
where the coefficient $A=5/2$ for $\Gamma=\gamma^t / \gamma^t\gamma_5$ cases \cite{LatticePartonLPC:2021gpi, LatticeParton:2022zqc, Yao:2022vtp} and $A=7/2$ for $\Gamma=\gamma^z / \gamma^z\gamma_5$ cases \cite{Holligan:2023rex, Yao:2022vtp}.
Since the fixed-order result contains large constants due to the renormalon divergence, which influences the perturbative convergence, we apply the leading renormalon resummation (LRR) method \cite{Zhang:2023bxs} to account for this effect. We then modify the Wilson coefficient as suggested in \cite{Zhang:2023bxs,Holligan:2023jqh}: 
\begin{align}
& C_0^{\mathrm{NLO+LRR}}(z, \mu)=C_0^{\mathrm{NLO}}(z, \mu) \nonumber\\
&\qquad +z \mu\left(C_{\mathrm{PV}}(z, \mu)-\sum_{i=0}^{k-1} \alpha_s^{i+1}(\mu) r_i\right),
\end{align}
where $r_i$ denote the coefficients of the renormalon series in $\alpha_s$ and $C_\mathrm{PV}(z,\mu)$ denotes the leading renormalon contribution for $C_0$ after a Borel transformation. The explicit forms of $r_i$ and $C_\mathrm{PV}$ can be found in Eqs.(12) and (13) of Ref.~\cite{Zhang:2023bxs}.

For a single lattice spacing $a$, Eq.(\ref{eq:matchingRenormalization}) can be written as
\begin{align}
	(m_0+\delta m)z-I_0=\ln\left[\frac{C_0^{\mathrm{NLO(+LRR)(\times RGR)}}\left(z, \mu\right)  }{ \tilde{M}^B(z, 0)}\right], \label{eq:fitformulaofm0}
\end{align}
where $I_0=\mathcal{I}^{\mathrm{lat}}\left(a^{-1}\right)$ is a constant, and the RGR improvement provides the scale conversion from lattice scale $z^{-1}$ to the renormalization scale $\mu$ in $\overline{\mathrm{MS}}$ scheme.

In the region $0\ll z<z_s$ , where perturbation theory is effective and discretization effects are not too large, we fit $m_0$
and $I_0$  based on Eq.(\ref{eq:fitformulaofm0}) over multiple ranges of $z$ up to a maximum of $z\simeq0.3$ fm. The fitting is carried out in the range $[z-a, z+a]$ for each $z$, using four different perturbative Wilson coefficient schemes: ``NLO” representing the fixed-order (NLO) $C_0$ at scale $\mu=m_D$,  ``NLO$\times$RGR" including the RGR improvement of $C_0$, and ``$\mathrm{(NLO+LRR)}\times\mathrm{RGR}$” combining both LRR and RGR improvements. The fit results with these different schemes are presented in the upper panel of Fig.\ref{fig:Ana_m0_dependence}, showing both statistical and systematic errors. The systematic errors are assessed by varying the lattice scale $\mu_0$ within a range from 0.8 to 1.2, accounting for the scaling uncertainties inherent in the lattice results.

\begin{figure}[htbp]
\centering
\includegraphics[scale=0.7]{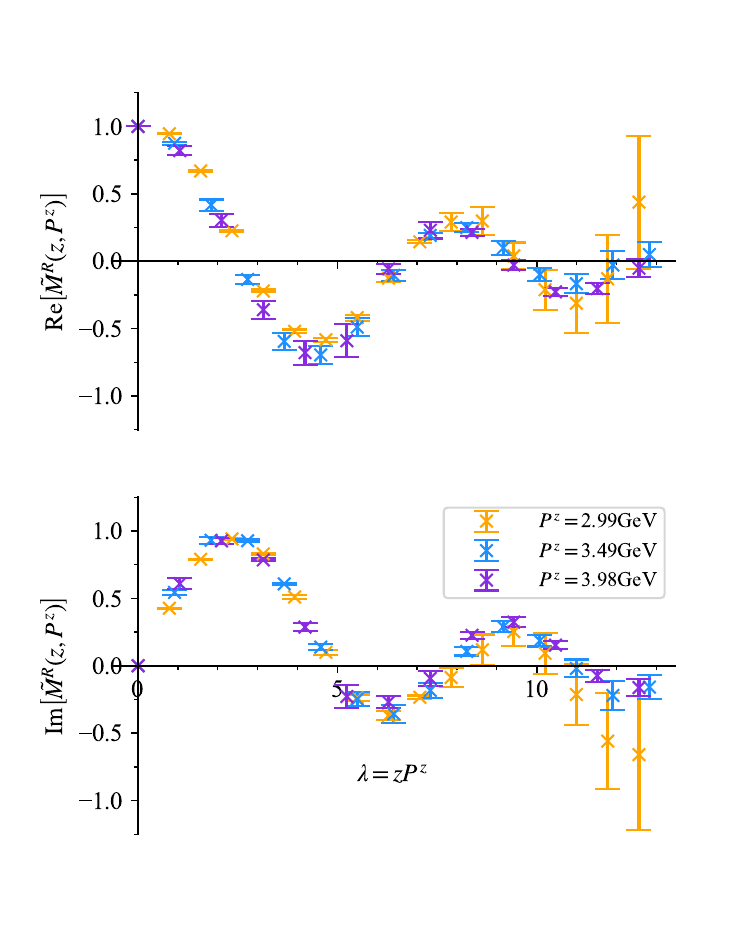}
\caption{The renormalized MEs of quasi DA $\tilde{M}^R(z, P^z)$ with boosted momenta $P^z=\{2.99,~3.49,~3.98\}$GeV. Both real and imaginary parts are shown as function of $\lambda=zP^z$. }
\label{fig:Ana_renormalized_ME}
\end{figure}

{Notice that the errors of ``NLO$\times$RGR" are significant larger than the ones of ``$\mathrm{(NLO+LRR)}\times\mathrm{RGR}$”, which mainly come from the systematic errors by varying the lattice scale $\mu_0$. This  reflects the systematic errors of fixed-order perturbation theory. Without performing the LRR improvement, the contributions from renormalons would be more significant, and contribute to a larger error after combining the variation of $\mu_0$.
When including the leading renormalon, the result  labeled as “(NLO+LRR)$\times$RGR” displays a clear plateau around $z\simeq0.2$ fm in the values of $m_0$, that indicates the LRR improvement can significantly reduce the dependence on $z$.}
We choose the result $m_0=(-0.173^{+0.014}_{-0.029})$GeV at $z=0.207$ fm, which exhibits  a clear plateau.  

We also investigate the scale dependence of $m_0$ to assess the impact of unaccounted higher-order terms in $C_0$. We compare the extracted results of $m_0$ from fixed-order $C_0^{\mathrm{NLO}}(z, \mu)$, as well as from the RGR improvement $C_0^{\mathrm{NLO\times RGR}}(z, \mu)$ and $C_0^{\mathrm{(NLO+LRR)}\times\mathrm{RGR}}(z, \mu)$. The results are shown in the lower panel of Fig.\ref{fig:Ana_m0_dependence}. One can see from the comparison that the RGR method significantly enhances the stability of $m_0$ after scale variation.

After removing the linear divergence and renormalon ambiguity, the renormalized MEs as a function of $\lambda=zP^z$ are depicted in Fig.~\ref{fig:Ana_renormalized_ME}. The MEs at different momenta exhibit  reasonable  consistency, indicating the saturation of equal-time correlations at large $P^z$. In showing the results, we choose $z_s=4a\simeq0.207$fm to ensure that the short distance correlations are reliably matched by perturbation theory.

\subsection{Extrapolating the long-range correlations}

Due to the finite volume of   gauge ensembles and the exponential increase in noise-to-signal ratio with spatial separation, the lattice results of nonlocal equal-time correlations tend to exhibit  large uncertainties  in the large-$\lambda$ region. To reconstruct the complete distributions,  we make the following assumption, drawing inspiration from the asymptotic behavior in the long-tail region $e^{-\lambda/\lambda_0}/|\lambda|^{d_1}$~\cite{Ji:2020brr,Gao:2021dbh} with the relation $\lambda= z\cdot P^z$.

More explicitly, we extrapolate the renormalized MEs   by using the following form \cite{Ji:2020brr}:
\begin{align}
	\tilde{M}^R(\lambda)=\left[\frac{c_1}{(-i\lambda_1)^{d_1}} + e^{i\lambda} \frac{c_2}{(i\lambda_2)^{d_2}} \right] e^{-\lambda/\lambda_0}, \label{eq:lambdaExtra}
\end{align}
where the parameterization inside the square brackets account for the algebraic behavior and is motivated by the Regge behavior \cite{Regge:1959mz} of the light-cone distributions in endpoint regions.

\begin{figure}[htbp]
\centering
\includegraphics[scale=0.7]{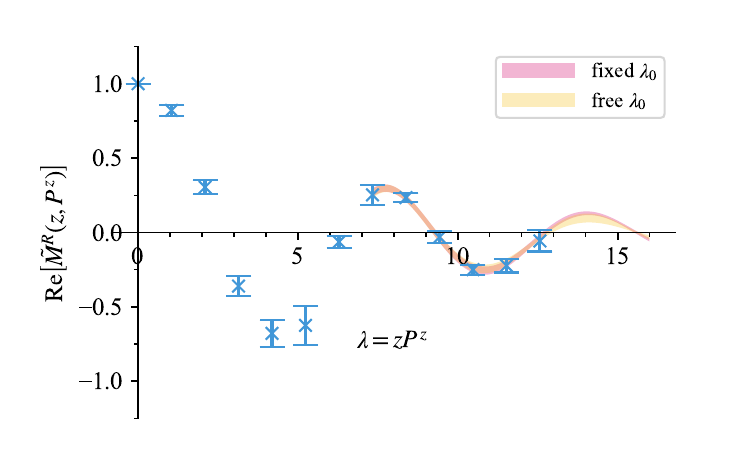}
\caption{The comparison of extrapolated results from fix the parameter $\lambda_0$ from the long-tail behavior of zero-momentum MEs, or treat it as a free parameter. We take the real part of $\tilde{M}^R(\lambda)$ at $P^z=3.98$GeV as example, and one can see that the extrapolated result from ``fixed $\lambda_0$'' is consistent with ``free $\lambda_0$'', while giving stricter restrictions of the errors.}
\label{fig:Ana_lambda_extra_with_fix_or_free_l0}
\end{figure}

As outlined earlier, the $\lambda_0$ can be obtained  by fitting the zero-momentum MEs, where the uncertainties are significantly smaller compared to those with higher momenta. This can provide a more stringent constraint on the fitting procedure. For illustration purpose, we conducted a comparison between two fitting strategies: one involving a ``fixed $\lambda_0$" determined by the result of $m_0$, and another strategy  using a ``free $\lambda_0$” parameter in the fit \cite{Gao:2021dbh}. The comparison of these two strategies is presented in Fig.~\ref{fig:Ana_lambda_extra_with_fix_or_free_l0}. It is evident from the figure that the extrapolated bands obtained from both strategies are in agreement, while the errors associated with the 	``fixed $\lambda_0$” approach are considerably smaller. To prevent underestimation of errors, we opt to employ the “free $\lambda_0$” strategy in our subsequent analysis.

\begin{figure}[htbp]
\centering
\includegraphics[scale=0.7]{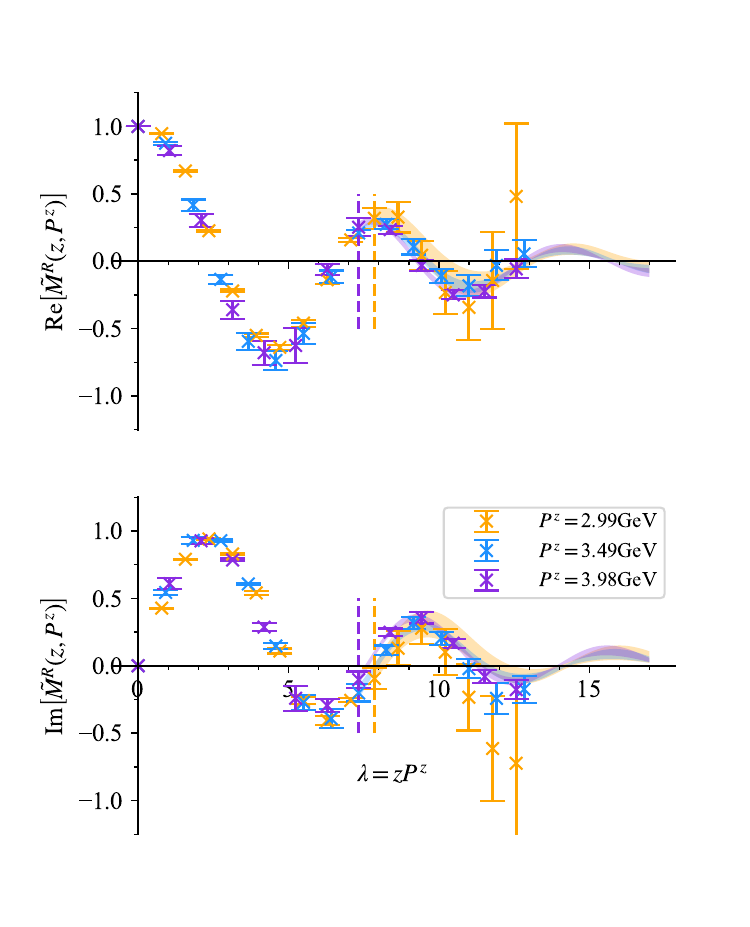}
\caption{Comparison of the original (data points) and extrapolated results (colored bands) at $P^z=\{2.99,~ 3.49,~ 3.98\}$GeV. We take the ``free $\lambda_0$'' strategy to perform the fit with fit range at $\lambda\geq\lambda_L$. The $\lambda_L$ for each $P^z$ are $\lambda_L=\{7.07,~ 7.33,~ 7.33\}$, as indicated by the dashed vertical lines. }
\label{fig:Ana_lambda_extrapolation}
\end{figure}

\begin{figure}[htbp]
\centering
\includegraphics[scale=0.7]{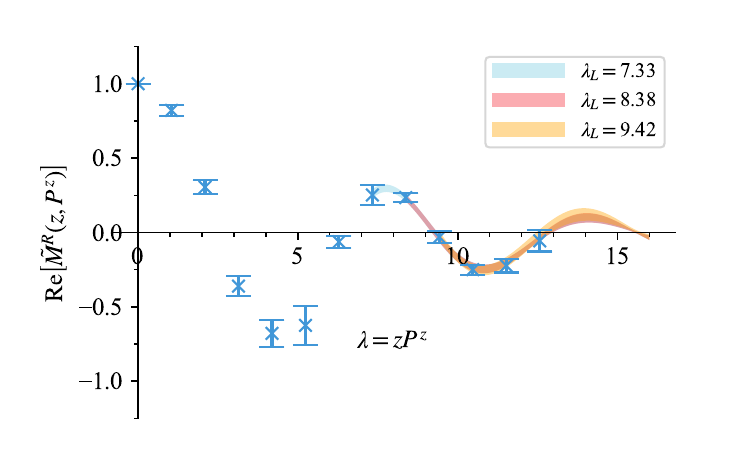}
\includegraphics[scale=0.7]{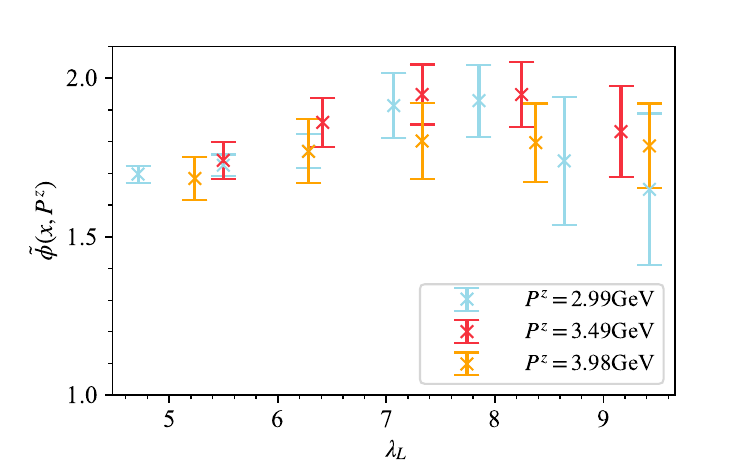}
\caption{Upper panel: Comparison of the $\lambda_L$ dependence of the real part of $\tilde{M}^R(z,P^z)$ at $P^z=3.98$GeV. The extrapolated results with $\lambda_L=\{7.33,~ 8.38,~ 9.42\}$ are consistent with each other, with only differences in errors. Lower panel: The $\lambda_L$ dependence of the quasi DAs $\tilde{\phi}(x, P^z)$ at fixed $x=0.25$ and $P^z=3.98$GeV, which come from the Fourier transformation of extrapolated MEs with different $\lambda_L$.}
\label{fig:Ana_lambda_extra_with_diff_lambdaL}
\end{figure}

The results before and after extrapolation are shown as the colored data points and bands in Fig.~\ref{fig:Ana_lambda_extrapolation}. We utilize the data at $\lambda\geq\lambda_L$ in the fitting procedure, as indicated by the data points to the right of the dashed line for each $P^z$ case. The choice of $\lambda_L$
 is crucial, as it should strike a balance between being too small or too large. On one hand, the region $\lambda\geq\lambda_L$ should capture the endpoint behavior of the momentum space distribution after Fourier transformation. On the other hand, the selection of $\lambda_L$  is constrained by the presence of rapidly increasing errors, necessitating a region where the MEs retain substantial non-zero values. To assess the impact of $\lambda_L$ on the extrapolation,  we compare the extrapolated results of the renormalized MEs at $P^z=3.98$  GeV obtained with different choices of $\lambda_L=\{7.33, 8.38, 9.42\}$ in the upper panel of Fig. \ref{fig:Ana_lambda_extra_with_diff_lambdaL}. In the lower panel, we contrast the results of the quasi DAs $\tilde{\phi}(x, P^z)$ in momentum space at a fixed momentum fraction of $x=0.25$ derived from the extrapolated data using varying values of $\lambda_L$.
It is evident that the results tend to stabilize after $\lambda_L\sim7$  for each momentum value, with only the errors showing an increasing trend. Based on this observation, we select $\lambda_L=\{7.07, 7.33, 7.33\}$ for the renormalized MEs at $P^z=\{2.99, 3.49, 3.98\}$  GeV as our final outcomes. To assess the systematic uncertainty stemming from the $\lambda$-extrapolation, we also explore the scenario where $\lambda_L$ is varied to $\{9.42, 9.16, 9.42\}$ and quantify the deviations between the results as the systematic uncertainty.

\begin{figure}[htbp]
\centering
\includegraphics[scale=0.7]{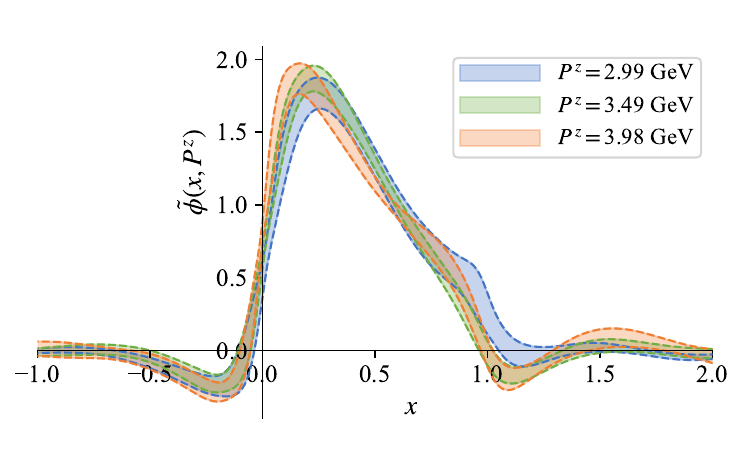}
\caption{Quasi DAs $\tilde{\phi}(x, P^z)$ with boosted momenta $P^z=\{2.99, 3.49, 3.98\}$GeV.}
\label{fig:Ana_quasi_DA_mom_space}
\end{figure}

After the extrapolation, we 
then Fourier transform the extrapolated MEs into momentum space to obtain the quasi DA $\tilde{\phi}(x, P^z)$ using
\begin{align}
	\tilde{\phi}(x, P^z) = \int_{-\infty}^{+\infty}\frac{dz}{2\pi}e^{-ixP^zz}\tilde{M}^R(z, P^z). 
\end{align}
The numerical results for $\tilde{\phi}(x, P^z)$  at various $P^z$ values are presented in Figure \ref{fig:Ana_quasi_DA_mom_space}. It is crucial to emphasize that the extrapolation formula given by Eq.(\ref{eq:lambdaExtra}) is formulated based on the behavior of the light-cone distribution at the endpoints as $x$ tends towards 0 and 1. This specifically pertains to the small-$x$ region, where $x$ is roughly less than $1/\lambda_L$.

\subsection{Matching the quasi DAs to QCD LCDAs} \label{sec:QuasiDAtoQCDLCDA}

Applying the matching relation given in Eq.~(\ref{eq:LaMETmatching}), we can establish the connection between the renormalized quasi DA $\tilde{\phi}(x, P^z)$ and the QCD LCDA $\phi(y,\mu)$ at next-to-leading order in $\alpha_s$.   
Note that the matching in Eq.(\ref{eq:LaMETmatching}) will suffer from large logarithmic terms at both $y \to 0$ (with the term $\sim \log[(2yP_z)^2/\mu^2]$) and $\bar{y} \to 0$ (with $\log[(2\bar{y}P_z)^2/\mu^2]$), where $y$ denotes the momentum fraction of the light quark and $\bar{y}$ denotes the momentum fraction of the heavy quark. Therefore, one needs to resum the large logarithms in these two regions separately.

Note that the counterterm in hybrid-ratio scheme does not contain any logarithm, and thus the $\mu$-dependence of the inverse matching coefficient $C^{-1}$ is same as the QCD LCDA $\phi(y,\mu)$ \footnote{The matrix form of perturbative matching coefficient $C$ can be formally expressed as $C=I+\frac{\alpha_sC_F}{2\pi}C^{(1)}$, where $I$ denotes a identity matrix. So its inverse can be expressed as $C^{-1}=I-\frac{\alpha_sC_F}{2\pi}C^{(1)}+\mathcal{O}(\alpha_s^2)$. }
, and the renormalized quasi DA is independent of $\mu$.  
Therefore, we resum the large logarithms by using the renormalization group equation of the inverse matching coefficient:
\begin{align}
	\frac{dC^{-1}(x, y, P^z, \mu)}{d\ln\mu^2} = \int_0^1 d\zeta V\left[\zeta, y,\alpha_s(\mu)\right]C^{-1}(x, \zeta, P^z, \mu), \label{eq:RGEofInverseC}
\end{align} 
which is same as the evolution of $\phi(y,\mu)$. $V\left[\zeta, y,\alpha_s(\mu)\right]$ is the Efremov-Radyushkin- Brodsky-Lepage  (ERBL) evolution kernel~\cite{Efremov:1979qk,Lepage:1980fj} of the QCD LCDA { which has been calculated up to three loops~\cite{Braun:2017cih}}. It is important to note that this evolution process is carried out for the momentum fraction $y$ within $\phi(y,\mu)$, as opposed to the momentum fraction $x$ within $\tilde{\phi}(x, P^z)$.

By solving Eq.(\ref{eq:RGEofInverseC}) 
we can perform the scheme conversion between the quasi DA, characterized by the scale $\mu_0=2yP^z$ or $2\bar{y}P^z$, to the QCD LCDA in the $\overline{\mathrm{MS}}$ scheme at scale $\mu$. In practice, we consider the RG resummation of the large logarithmic terms in the two end-point regions separately. In the region $y\to0$, we choose $\mu_0 = 2yP^z$ and resum the terms $\sim \log\left[(2yP^z)^2/\mu^2\right]$, and in the region $y\to1$, we choose $\mu_0=2\bar{y}P^z$ and resum the terms $\log\left[(2\bar{y}P^z)^2/\mu^2\right]$. The $\overline{\mathrm{MS}}$ scale $\mu$ is chosen as $\mu=m_D$ thus combining the factorization scale of quasi DA, and the renormalization scale of QCD LCDA.
Given the uncertainty associated with the initial resummation scale, we systematically vary the scale $\mu_0$ to assess the robustness of the perturbative matching procedure and to estimate the pertinent systematic uncertainties, following a similar approach detailed in Sec.~\ref{sec:RenormInRatioScheme}. The selection of different resummation scales captures higher-order effects in $\alpha_s$, which are expected to vanish when the perturbative calculation is extended to all orders.

\begin{figure}[htbp]
\centering
\includegraphics[scale=0.7]{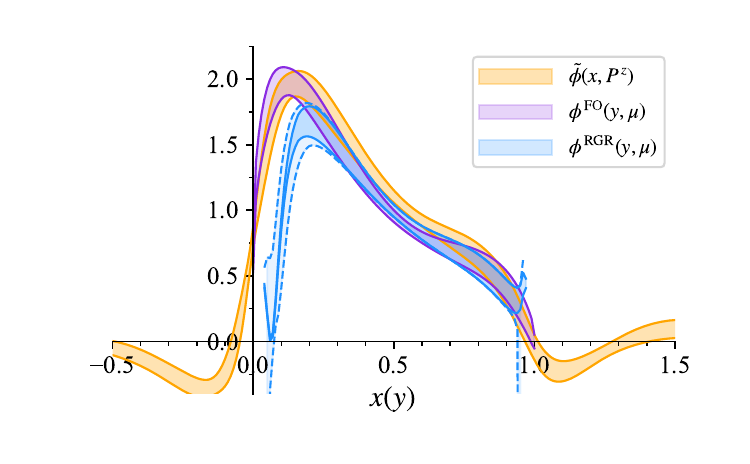}
\caption{Quasi DA $\tilde{\phi}(x,P^z)$ with $P^z=3.98$GeV, and matched QCD LCDAs from NLO kernel with fixed-order perturbation theory $\phi^{\mathrm{FO}}(y,\mu)$ and RGR improvement $\phi^{\mathrm{RGR}}(y,\mu)$. The scale $\mu$ is chosen as $m_D$. The first two results only contain statistical errors, while the latter one not only contains statistical error (solid lines in the blue band), but also the systematic error arising from the scale variation (dashed lines).}
\label{fig:Ana_QCDLCDA_compare_Quasi_RGR_FO}
\end{figure}

Fig.~\ref{fig:Ana_QCDLCDA_compare_Quasi_RGR_FO} presents a comparison between the quasi distribution amplitude $\tilde{\phi}(x,P^z)$ at $P^z=3.98$ GeV, and the QCD LCDAs obtained from matching with fixed-order perturbation theory, denoted as $\phi^{\mathrm{FO}}(y,\mu)$, and with the RGR improvement, denoted as $\phi^{\mathrm{RGR}}(y,\mu)$. The RGR improvement is performed on both endpoint regions at $y\to0$ and $y\to1$. The dashed lines in $\phi^{\mathrm{RGR}}(y,\mu)$  represent the variation of the resummation scale from $0.8\mu_0$
to $1.2\mu_0$, corresponding to approximately a $20\%$ change in the strong coupling constant $\alpha_s(\mu_0)$
around $y\sim0.2$. In the intermediate $y$ region, where the physical scale of the quasi distribution amplitude $\mu_0$
is in proximity to the $\overline{\mathrm{MS}}$  scale $\mu=m_D$  and the resummed logarithms are not excessively large, the RGR-improved and fixed-order matched QCD LCDAs exhibit a high degree of consistency.

In the small $y$ region, such as $0.1\lesssim y\lesssim0.3$, a significant discrepancy between the two distributions is apparent, indicating the increasing importance of the resummation effect which enhances the accuracy of the theoretical predictions. Additionally, at the endpoint where $y\lesssim0.1$ and $y\gtrsim0.9$, the sharp change in behavior observed in $\phi^{\mathrm{RGR}}(y,\mu)$  is attributed to the emergence of the Landau pole, signifying the breakdown of perturbative matching in this region.

Moreover, it is important to note that the resummation of logarithmic terms from higher orders can disrupt the normalization of the QCD LCDAs at fixed order. Specifically, we observe that $\int_0^1dy \phi^{\mathrm{FO}}(y,\mu)=1\neq\int_0^1dy \phi^{\mathrm{RGR}}(y,\mu)$ in Fig. \ref{fig:Ana_QCDLCDA_compare_Quasi_RGR_FO}. For example, we have determined that the central values of the QCD LCDAs obtained from the RGR-improved matching kernel are typically normalized to around 0.81. A similar normalization issue also exists in the analysis of QCD LCDAs from the matching of HQET LCDAs~\cite{Beneke:2018wjp}. Following this approach, we rescale the QCD LCDA $\phi^{\mathrm{RGR}}(y,\mu)$ to ensure its normalization to 1 for subsequent phenomenological analyses. A similar normalization problem also exists in the analysis of QCD LCDAs from the matching of HQET LCDAs~\cite{Beneke:2018wjp}, following it, we rescale the QCD LCDA $\phi^{\mathrm{RGR}}(y,\mu)$ to ensure its normalization to 1 for subsequent phenomenological analyses.
To address the issue of improper normalization, one potential approach is to calculate the lowest moments of the QCD LCDAs for charmed mesons directly. This method is complementary to the LaMET approach \cite{Ji:2022ezo} and  facilitates a straightforward comparison of these findings, potentially leading to valuable insights.

\begin{figure}[htbp]
\centering
\includegraphics[scale=0.7]{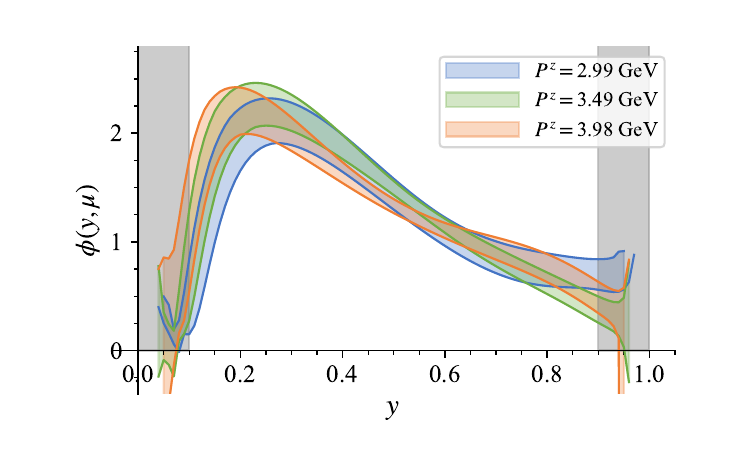}
\caption{The $P^z$ dependence of the QCD LCDAs. Only statistic error and systematic error from scale variation in RGR are included in the error bands. The grey band indicates the region ($y\lesssim\Lambda_{\mathrm{QCD}}/P^z\simeq0.1$) where the matching in LaMET failed.}
\label{fig:Ana_LCDA_with_mom}
\end{figure}

In Fig. \ref{fig:Ana_LCDA_with_mom}, we present the $P^z$-dependence of the QCD LCDA with RGR-improved NLO matching. The momenta of the boosted heavy meson are selected as $P^z=\{6,~ 7,~ 8\}\times2\pi/L\simeq\{2.99,~ 3.49,~ 3.98\}$ GeV. Our analysis reveals that the results at different $P^z$ values exhibit consistent behavior within the margins of error, suggesting a potential saturation at large momenta which needs more detailed analysis in the future. Given the consistency across various momenta, we opt to utilize the result obtained at $P^z=3.98$ GeV as our final prediction for the QCD LCDA
, and compute the systematic error arising from power corrections at finite $P^z$ by comparing the central value of the result at $P^z=3.98$GeV with that at $P^z=2.99$GeV.

\subsection{Determination of HQET LCDA} \label{sec:DeterminHQETLCDA}

\begin{figure}[htbp]
\centering
\includegraphics[scale=0.7]{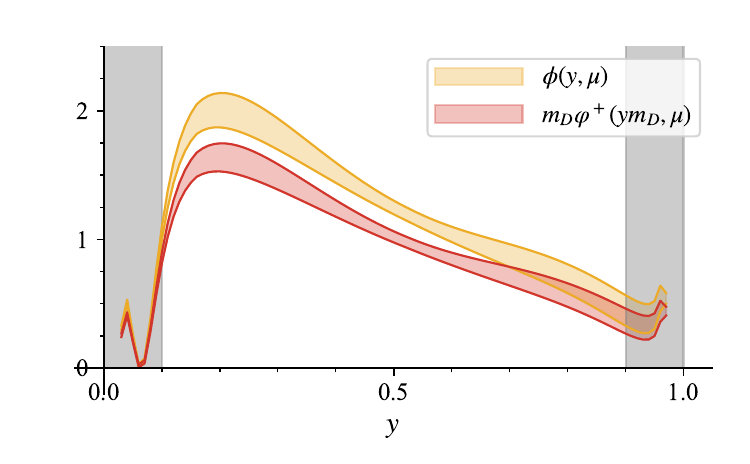}
\caption{Comparison of QCD LCDA and HQET LCDA obtained from matching via Eq.(\ref{eq:peak_varphi_+}) at $\mu=m_D$. Only statistic errors are obtained in both results. It should be noted that although the full distribution of HQET LCDA is shown, only the part satisfies $\omega\sim\Lambda_{\mathrm{QCD}}$ is valid.}
\label{fig:Ana_LCDA_compare_QCD_HQET}
\end{figure}

As discussed in Sec.~\ref{sec:HQETmatching}, we extract the peak region of the HQET LCDA from the QCD LCDA using a multiplicative factorization approach. The resulting matched HQET LCDA is depicted by the red band in Fig. \ref{fig:Ana_LCDA_compare_QCD_HQET}. From this figure, one can observe that the peak position of the QCD LCDA $\phi(y, \mu=m_D)$  (orange band) is situated within the range $y\in[0.2, 0.3]$, corresponding to the scenario where the light quark carries a typical light-cone momentum fraction $y\sim\Lambda_{\mathrm{QCD}}/m_D$. 
Furthermore, at very large scales $\mu\gg m_D$, the QCD LCDA reverts back to its asymptotic form akin to that of a light meson.

Before reaching a final conclusion regarding the peak region, we address potential sources of systematic errors in our analysis, which may arise from
\begin{itemize}
	\item the extrapolation of long-range correlations. We use the data with $\lambda\geq\lambda_L$ for  the fit, and vary the values of $\lambda_L$ from $\{7.07,~ 7.33,~ 7.33\}$ to $\{9.42,~ 9.16,~ 9.42\}$ (for the renormalized MEs at $P^z=\{2.99, 3.49, 3.98\}$GeV) to estimate the systematic error. 
	\item the scale uncertainty of the MEs renormalized in hybrid-ratio scheme. In the process of matching from lattice results to the light-cone quantities in the $\overline{\mathrm{MS}}$  scheme, uncertainties in the initial conditions can introduce systematic errors. In the renormalization group running of the Wilson coefficient discussed in Sec. \ref{sec:RenormInRatioScheme}, we adopt an initial condition for scale evolution of $c'\times2e^{-\gamma_E}z^{-1}$, while in the renormalization group running of the perturbative matching kernel as discussed in Sec. \ref{sec:QuasiDAtoQCDLCDA}, we use $c'\times2xP^z$. To assess the impact of these initial conditions, we vary the factor $c'$  in the range of 0.8 to 1.2 to estimate the systematic error.
	\item the difference between QCD LCDAs extracted at different large $P^z$. Rather than extrapolating $P^z$
  to infinite momentum, we adopt the QCD LCDA result at $P^z=3.98$
  GeV as our final prediction. To quantify the systematic error, we calculate the difference between this prediction and the one at $P^z=2.99$
  GeV.
\end{itemize}

\begin{figure}[htbp]
\centering
\includegraphics[scale=0.7]{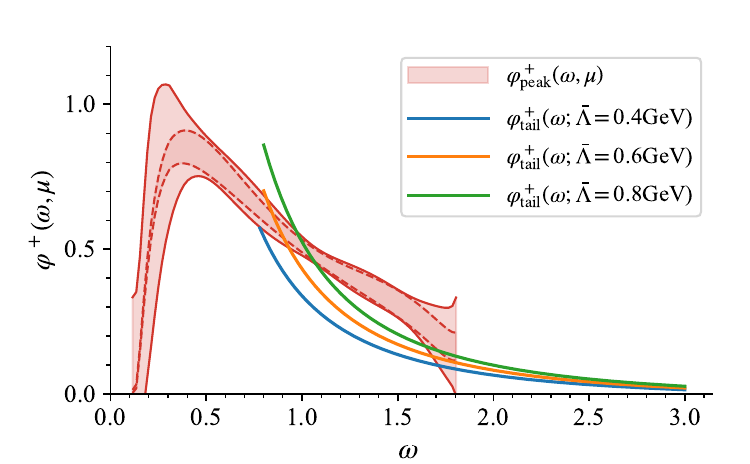}
\caption{The result of HQET LCDA in the peak region $\varphi_{\mathrm{peak}}^+(\omega,\mu)$ and in the tail region $\varphi_{\mathrm{tail}}^+(\omega; \bar{\Lambda})$. The dashed lines in $\varphi_{\mathrm{peak}}^+(\omega,\mu)$ indicate the upper and lower limits of the result containing only statistical error, while the solid lines indicate the result including both statistical and systematic errors. Three colored solid lines of $\varphi_{\mathrm{tail}}^+(\omega; \bar{\Lambda})$ represent the results at $\bar{\Lambda}=\{0.4, 0.6, 0.8\}$GeV, where $\bar{\Lambda}$ denotes the power corrections in the perturbative calculations.}
\label{fig:Ana_HQETLCDA_compare_peak_tail}
\end{figure}

Combining the statistical and systematic errors discussed earlier, we derive the numerical results for the heavy meson HQET LCDA from lattice QCD through a two-step matching process, which contributes to the peak region of the final prediction. The results are shown in Fig.\ref{fig:Ana_HQETLCDA_compare_peak_tail}, in which the band represented by dashed lines corresponds to the result including only statistical errors, while the one with solid lines represents the result incorporating both statistical and systematic errors.

The tail region of HQET LCDA contains only hard-collinear contributions, which would contribute only through power corrections. We employ the one-loop level result from Ref.~\cite{Lee:2005gza} that is given in Eq.~(\ref{eq:tail_varphi_+}).  
The parameter $\bar{\Lambda}$ reflects the power correction and usually be chosen at hundreds of MeV. Here we use $\bar{\Lambda}=\{0.4,\, 0.6,\, 0.8\}$GeV to estimate the power correction in Eq.(\ref{eq:tail_varphi_+}), shown as the colored lines in Fig.\ref{fig:Ana_HQETLCDA_compare_peak_tail}.

The power counting of parameters in HQET LCDA reveals that the momentum of the light quark in the peak region satisfies $\omega\sim\mathcal{O}(\Lambda_{\mathrm{QCD}})$, while in the tail region it satisfies $\omega\sim\mathcal{O}(m_D)$. This suggests that the boundary between them lies in the range $\Lambda_{\mathrm{QCD}}\ll\omega_{b}\ll m_D$. From Fig. \ref{fig:Ana_HQETLCDA_compare_peak_tail}, we observe that the intersection of the peak and tail regions occurs around 0.9 GeV, confirming the aforementioned statement. Consequently, we merge them into a continuous distribution with
\begin{align}
	\varphi^+(\omega,\mu)=\varphi^+_{\mathrm{peak}}(\omega,\mu)\theta(\omega_{b}-\omega) + \varphi^+_{\mathrm{tail}}(\omega,\mu)\theta(\omega-\omega_b),
\end{align}
where the intersection position $\omega_b=\{0.79,~ 0.86,~ 0.94\}$GeV for $\bar{\Lambda}=\{0.4,~0.6,~0.8\}$GeV. In order to ensure a smooth transition between the peak and tail regions, we utilize a polynomial filter, specifically the Savitzky-Golay filter, on the results within a vicinity of $\delta=0.05$ GeV around $\omega_b$
  to smoothen the curve.

 \begin{figure}[htbp]
\centering
\includegraphics[scale=0.7]{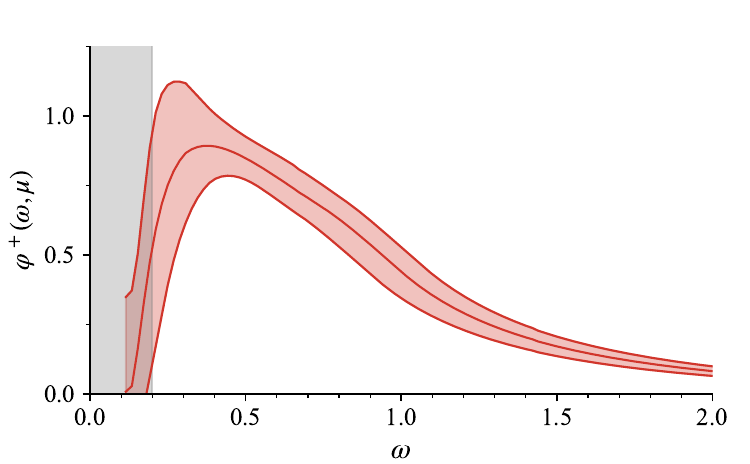}
\caption{Final result of HQET LCDA $\varphi^+(\omega,\mu)$ at $\mu=m_D$. All statistical and systematic errors are included. The grey band indicates the region where our result suffers significant power corrections, approximately in $\omega\lesssim0.19$GeV.}
\label{fig:Ana_final_HQETLCDA}
\end{figure}

The combined result of the peak and tail regions is illustrated in Fig.\,$ \ref{fig:Ana_final_HQETLCDA}$. The peak region is derived from lattice QCD calculations and encompasses both statistical and systematic errors, while the tail region is determined from one-loop perturbation theory and includes solely systematic errors. The grey band highlights the range where our findings are notably affected by power corrections, specifically around $\omega\lesssim0.19$ GeV.


\section{Phenomenological discussions}

\subsection{Tabulated  results  for HQET LCDA}
 
We present the numerical results of the peak region of HQET LCDA $\varphi^+_{\mathrm{peak}}(\omega,\mu)$ in Tab.\ref{tab:binned_results}, which  are determined from lattice QCD calculations. 
As mentioned earlier, in the small $\omega$ region ($\omega\sim m_D\Lambda_{\mathrm{QCD}}/P^z$), the results are affected by significant power corrections, while in the large $\omega$ region ($\omega\gg\Lambda_{\mathrm{QCD}}$), the HQET LCDA prediction relies on perturbative calculations. Therefore, we present the numerical results for the HQET LCDA $\varphi^+(\omega,\mu)$  in the peak region with $\omega\in[0.192,~0.806]$ GeV. These results include the central values $\overline{\varphi^+_{\mathrm{peak}}}(\omega, \mu)$ and the covariance matrix $\mathrm{Cov}(\varphi^+_{\mathrm{peak}}(\omega, \mu))$ associated with different $\omega$ values. The renormalization scale $\mu$ is chosen to be the charmed meson mass $m_D$.

\begin{table*}[ht]
  \caption{Numerical results for the HQET LCDA $\varphi^+(\omega,\mu)$ at ranging $\omega\in[0.192,~0.806]$GeV, including the central value $\overline { \varphi^+_{\mathrm{peak}}(\omega, \mu)}$ and covariant matrix $\mathrm{Cov}(\varphi^+_{\mathrm{peak}}(\omega, \mu))$ (in units of $10^{-3}$). The non-diagonal terms in the covariant matrix come from the correlations between different gauge ensembles. The renormalization scale $\mu$ is chosen to be $m_D$. }
  \renewcommand{\arraystretch}{2.0}
\setlength{\tabcolsep}{2.5mm}
\centering
\begin{tabular}{cccccccccc}
    \hline\hline
    $\omega$ (GeV) & 0.192 & 0.269 & 0.346 & 0.422 & 0.499 & 0.576 & 0.653 & 0.73 & 0.806 \\
    \hline
    $\overline { \varphi^+_{\mathrm{peak}}(\omega, \mu)}$ & 0.474 & 0.802 & 0.888 & 0.885 & 0.848 & 0.797 & 0.740 & 0.682 & 0.627 \\
    \hline
     & 169.476 & 1.068 & 1.269 & 1.295 & 1.213 & 1.057 & 0.854 & 0.627 & 0.392 \\
     & 1.068 & 103.227 & 2.603 & 2.727 & 2.579 & 2.244 & 1.794 & 1.289 & 0.774 \\
     & 1.269 & 2.603 & 33.617 & 3.634 & 3.474 & 3.036 & 2.426 & 1.734 & 1.032 \\
    $\mathrm{Cov}(\varphi^+_{\mathrm{peak}}(\omega, \mu))$  & 1.295 & 2.727 & 3.634 & 10.583 & 3.819 & 3.371 & 2.724 & 1.979 & 1.213 \\
    ~~(in units of $10^{-3}$)   &  1.213 & 2.579 & 3.474 & 3.819 & 6.014 & 3.349 & 2.759 & 2.062 & 1.330 \\
    & 1.057 & 2.244 & 3.036 & 3.371 & 3.349 & 6.117 & 2.589 & 2.009 & 1.379 \\
    & 0.854 & 1.794 & 2.426 & 2.724 & 2.759 & 2.589 & 7.011 & 1.843 & 1.357 \\
    & 0.627 & 1.289 & 1.734 & 1.979 & 2.062 & 2.009 & 1.843 & 7.118 & 1.263 \\
    & 0.392 & 0.774 & 1.032 & 1.213 & 1.330 & 1.379 & 1.357 & 1.263 & 6.120 \\
    \hline
\end{tabular}
\label{tab:binned_results}
\end{table*}

\subsection{Comparison with phenomenological models}

\begin{figure}[!t]
  \centering
  \includegraphics[scale=0.7]{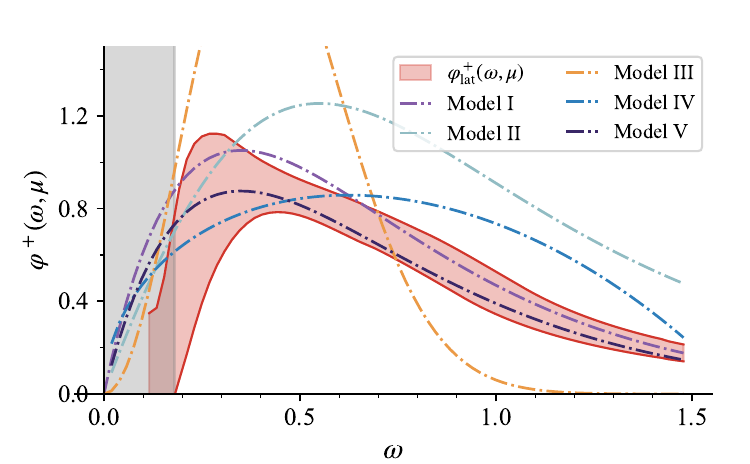}
  \caption{Comparison of the numerical results of $\varphi^{+}\left(\omega, \mu\right)$ from lattice QCD with several commonly used phenomenological models in Eq.(\ref{eq:models}). The renormalization scale $\mu$ of our result is chosen to be $m_D$.  We only show the central values of the models as colored dashed lines, and interpret their spread as error estimate.}
  \label{fig:comparetopheno}
  \end{figure}
  
  \begin{table*}[!t]
    \caption{Our results of the parameters by fitting the models in Eq.(\ref{eq:models}) compare to those values given in references \cite{Wang:2015vgv,Beneke:2018wjp,Gao:2021sav}. Due to the unique behavior of model III, we are unable to obtain a good fit using this model ($\chi^2/$d.o.f$>$3 and the fitted results are unable to describe the data). When fitting model V, we adopt the values for parameters $\alpha$ and $\beta$ from Ref.~\cite{Beneke:2018wjp,Gao:2021sav}. }
    \renewcommand{\arraystretch}{2.0}
    \setlength{\tabcolsep}{2.5mm}
    \begin{tabular}{c c c c c c }
    \hline\hline
       Models &  I & II  & III & IV  & V     \\\hline
       This work &  $\omega_0=437\pm59$MeV  &  $\omega_0=641\pm134$MeV   &   ---------     &  $\omega_0=374\pm35$MeV & $\omega_0=432\pm52$MeV  \\
         &  & $\sigma_B^{(1)}=2.70\pm1.86$GeV &  &  &   \\
      References & $\omega_0=354_{-30}^{+38}$MeV & $\omega_0=368_{-32}^{+42}$MeV & $\omega_0=389_{-28}^{+35}$MeV & $\omega_0=303_{-26}^{+35}$MeV  & $\omega_0=350$MeV \\
       &  & $\sigma_B^{(1)}=1.4\pm0.4$GeV &  &  &   \\
       \hline
    \end{tabular} 
    \label{tab:fit_model}
 \end{table*}

In this subsection, we will  compare our results for the HQET LCDA with those from previous studies based on various phenomenological models \cite{Grozin:1996pq,Braun:2003wx,Beneke:2018wjp}:
\begin{align}
&\varphi_{\mathrm{I}}^{+}\left(\omega, \mu_0\right)=\frac{\omega}{\omega_0^2} e^{-\omega / \omega_0} , \nn\\
&\varphi_{\mathrm{II}}^{+}\left(\omega, \mu_0\right)=\frac{4}{\pi \omega_0} \frac{k}{k^2+1}\left[\frac{1}{k^2+1}-\frac{2\left(\sigma_B^{(1)}-1\right)}{\pi^2} \ln k\right] , \nn\\
&\varphi_{\mathrm{III}}^{+}\left(\omega, \mu_0\right)=\frac{2 \omega^2}{\omega_0 \omega_1^2} e^{-\left(\omega / \omega_1\right)^2} , \nn\\
&\varphi_{\mathrm{IV}}^{+}\left(\omega, \mu_0\right)=\frac{\omega}{\omega_0 \omega_2} \frac{\omega_2-\omega}{\sqrt{\omega\left(2 \omega_2-\omega\right)}} \theta\left(\omega_2-\omega\right) , \nn\\
&\varphi_{\mathrm{V}}^+(\omega, \mu_0) = \frac{\Gamma(\beta)}{\Gamma(\alpha)} \frac{\omega}{\omega_0^2} e^{-\omega/\omega_0} U(\beta - \alpha, 3-\alpha, \omega/\omega_0).
\label{eq:models}
\end{align}
The parameters of the first four models are collected as follows \cite{Wang:2015vgv}: in model I, $\omega_0=354_{-30}^{+38}$MeV; in model II, $\omega_0=368_{-32}^{+42}$MeV, $\sigma_B^{(1)}=1.4\pm0.4$GeV and $k={\omega}/{1\,\mathrm{GeV}}$; in model III, $\omega_0=389_{-28}^{+35}$MeV and $\omega_1={2 \omega_0}/{\sqrt{\pi}}$; in model IV, $\omega_0=303_{-26}^{+35}$MeV and $\omega_2={4 \omega_0}/{(4-\pi)}$. These parameters are all given at $\mu=1$GeV. In model V, $U(a,b,z)$ denotes the second kind confluent hypergeometric function, and the parameters $\omega_0$, $\alpha$ and $\beta$ can be solved from the numerical results of $\lambda_B$ and $\hat{\sigma}_1$ \cite{Beneke:2018wjp,Gao:2021sav}. 

In Fig.~\ref{fig:comparetopheno}, we compare our results with those parametrized in Eq.~(\ref{eq:models}). To avoid cluttering the plot and due to the challenges in estimating uncertainties in model-based analyses, we only display the central values as colored dashed lines. The spread of these lines serves as an estimate of systematic  errors in the models, and we refrain from discussing the relative merits of these different models. It is worth noting that while our results are  consistent with the combined phenomenological model estimates, they exhibit  better theoretical foundation.  The analysis in this study was primarily intended as a demonstration. We are confident that future lattice simulations following similar methodologies will significantly improve precision, possibly eliminating the necessity for model-based analyses.

Furthermore, we have conducted data fitting based on the models outlined in Eq.(\ref{eq:models}), and the corresponding results are detailed in Table \ref{tab:fit_model}.  The fit range is selected based on the range that yields the optimal quality of fit to the data.  Due to the unique behavior of model III, we are unable to obtain a good fit using this model ($\chi^2/$d.o.f$>$2 and the fitted results are unable to describe the data). When fitting model V, we adopt the parameter values for $\alpha$ and $\beta$ as specified in Ref.~\cite{Beneke:2018wjp,Gao:2021sav}, which are not significantly influenced by the data.  Notably, in these models, the parameter $\omega_0$ is equivalent to $\lambda_B$.

\subsection{Determination of the inverse and inverse-logarithmic moments} 

The first inverse moment $\lambda_B^{-1}$ and inverse-logarithmic moments $\sigma_B^{(n)}$ play a pivotal role in QCD factorization theorems and lightcone sum rule studies in heavy flavor physics. These quantities are defined as:
\begin{align}
  \lambda_B^{-1}(\mu)=&\int_0^{\infty} \frac{d\omega}{\omega}  \varphi^+(\omega,\mu) , \label{eq:lambdaB_integral} \\
  \sigma^{(n)}_B(\mu) =& \lambda_B(\mu)\int_0^\infty \frac{d\omega}{\omega} \ln \left(\frac{\mu}{\omega}\right)^{(n)}\varphi^+(\omega,\mu). \label{eq:sigmaB_integral}
 \end{align} 
While the first inverse moment $\lambda_B^{-1}$ and the first inverse-logarithmic moments $\sigma_B^{(1)}$ have been estimated in various approaches~\cite{Belle:2018jqd,Khodjamirian:2020hob,Lee:2005gza,Braun:2003wx,Grozin:1996pq}, there is considerable scope for enhancing their reliability and precision. In the case of $\sigma^{(2)}_B$ and subsequent orders, no existing results are currently available.

\begin{figure}[htbp]
  \centering
  \includegraphics[scale=0.7]{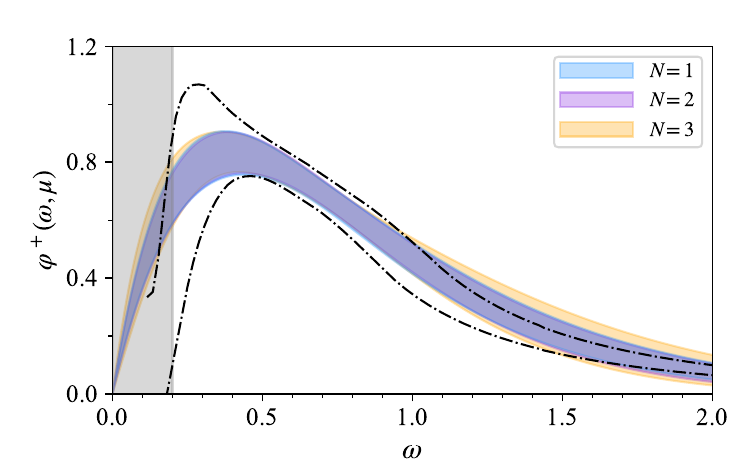}
  \includegraphics[scale=0.7]{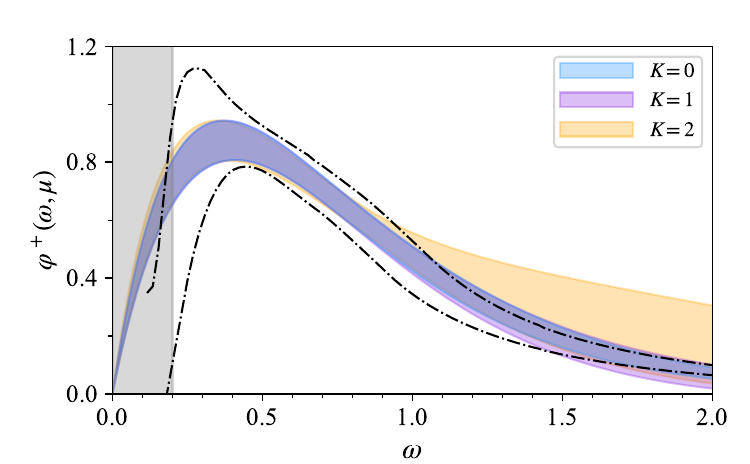}
  \caption{Fit results based on the parametrizations in Eq. (\ref{eq:HQETLCDAsmallomegaexpansion}) up to the $N$-th order (upper panel) and in Eq. (\ref{eq:HQETLCDAsmallomegaexpansion2}) up to the $K$-th order (lower panel). The fitting ranges adopted are $\omega\in[0.192, 0.806]$GeV. The dashed lines indicate the upper and lower limits of the original data. }
  \label{fig:omega_expansion}
\end{figure}

To determine these values from the HQET LCDA we have developed, it is essential to integrate the HQET LCDA $\varphi^+(\omega,\mu)$ over $\omega$ from 0 to infinity. As noted earlier, our results will be influenced by significant power corrections at low $\omega$, presenting difficulties for accurate predictions in this regime.
Alternatively, we can employ some model-independent parametrization formulas to extrapolate our results towards the region near $\omega=0$. 
One possible parameterization is:
\begin{align}
\varphi^{+}&\left(\omega, \mu\right)=\sum_{n=1}^{N} c_n\frac{\omega^n}{\omega_0^{n+1}} e^{-\omega / \omega_0} \nn\\
=&\frac{c_1\omega}{\omega_0^2} \left[ 1 + c_2'\frac{\omega}{\omega_0} + c_3'\left(\frac{\omega}{\omega_0}\right)^2 + \cdots \right] e^{-\omega / \omega_0},
\label{eq:HQETLCDAsmallomegaexpansion}
\end{align}
with $c_n'\equiv c_n/c_1$. One benefit of this parametrization is that for small $\omega\ll \omega_0$, higher moments become less influential. By expanding Eq.(\ref{eq:HQETLCDAsmallomegaexpansion}) up to the first ($N=1$), second ($N=2$), and third ($N=3$) order terms, we can determine the parameters as:
\begin{align}
  N=1:~ & \omega_0=0.388 (46),~ c_1=0.916 (81); \nn\\
  N=2:~ & \omega_0=0.340 (80),~ c_1=0.68 (36), \nn\\ & c_2'=0.17 (33); \nn\\
  N=3:~ & \omega_0=0.32(15),~ c_1=0.63(44), \nn\\ & c_2'=0.13 (38), ~ c_3'=0.03(19).
\end{align}
The outcomes derived using the aforementioned parameters are illustrated in the upper panel of Fig.~\ref{fig:omega_expansion}. It is evident that the results obtained from expanding to the $N$-th order ($N=1,~2,~3$) exhibit consistency, signifying the favorable convergence of the expansion depicted in Eq.(\ref{eq:HQETLCDAsmallomegaexpansion}).

An alternative model-independent parametrization is based on an expansion in generalized Laguerre polynomials, as provided by Ref. \cite{Feldmann:2022uok}:
\begin{align}
	\varphi^+(\omega,\mu)= \frac{\omega e^{-\omega / \omega_0}}{\omega_0^2} \sum_{k=0}^K \frac{a_k(\mu)}{1+k}L_k^{(1)}(2\omega/\omega_0), \label{eq:HQETLCDAsmallomegaexpansion2}
\end{align}
where $L_k^{(1)}$ denote the associated Laguerre polynomials. 
Note that the parameterizations given by Eq.(\ref{eq:HQETLCDAsmallomegaexpansion}) and Eq.(\ref{eq:HQETLCDAsmallomegaexpansion2}) are both complete, with the latter having the advantage of being an expansion based on a complete set of orthogonal polynomials. 
By expanding Eq.(\ref{eq:HQETLCDAsmallomegaexpansion2}) up to the first ($K=0$), second ($K=1$), and third ($K=2$) order terms, we can determine the parameters as:
\begin{align}
	K=0:~ & \omega_0=0.389 (29), ~ a_0=0.925 (62); \nn\\
  	K=1:~ & \omega_0=0.446 (95), ~ a_0=1.05 (19), \nn\\ & a_1=0.15 (26); \nn\\
  	K=2:~ & \omega_0=0.47 (10), ~ a_0=1.14 (21), \nn\\ & a_1=0.17 (24), ~ a_2=0.16 (18).
\end{align}
 The curves obtained based on the fitting results are shown in the lower panel of Fig.~\ref{fig:omega_expansion}. From the comparison of the curves obtained from the two parameterizations, it is evident that the results are basically consistent, with only the error in the tail region differing.  This highlights the difference in the convergence of the two parameterization schemes.

Using the extrapolated data, we can evaluate the integrals in Eqs.(\ref{eq:lambdaB_integral}-\ref{eq:sigmaB_integral}) to determine the numerical values of the first inverse moment $\lambda_B^{-1}$ and the first two inverse-logarithmic moments $\sigma_B^{(1,2)}$ of the HQET LCDA. 
These results are displayed in the upper panel of Table \ref{tab:model_independent_moments}, where $N$ and $K$ denote the truncation order in the expansion of Eq.(\ref{eq:HQETLCDAsmallomegaexpansion}) and Eq.(\ref{eq:HQETLCDAsmallomegaexpansion2}) respecticvely. Note that these results are obtained at $\mu=m_D$ through fitting the data in Tab.\ref{tab:binned_results}. 
To evolve them to the soft scale $\mu=1$ GeV, which is commonly used in phenomenology, we adopt the following RG evolution equations  \cite{Bell:2008er,Bell:2013tfa,Wang:2015vgv}
\begin{align}
    \frac{\lambda_B\left(\mu_0\right)}{\lambda_B(\mu)} = & 1+\frac{a_s(\mu_0)}{4} \ln \frac{\mu}{\mu_0}\left[2-2 \ln \frac{\mu}{\mu_0}-4 \sigma_B^{(1)}\left(\mu_0\right)\right] \nn\\
    &\qquad  +\mathcal{O}\left(a_s^2\right), \\
    \sigma_B^{(1)}(\mu) = & \sigma_B^{(1)}\left(\mu_0\right)+\ln \frac{\mu}{\mu_0} \bigg(1+a_s(\mu_0) \nn\\
    & \qquad \times\left[\left(\sigma_B^{(1)}\left(\mu_0\right)\right)^2-\sigma_B^{(2)}\left(\mu_0\right)\right]\bigg)+\mathcal{O}\left(a_s^2\right),
\end{align}
where $a_s(\mu)\equiv\alpha_s(\mu)C_F/\pi$. The results of $\lambda_B$ and $\sigma_B^{(1)}$ after RG evolution are collected in the central panel of Table \ref{tab:model_independent_moments}. The values of $\sigma_B^{(2)}$ are not included because their evolution depend on the value of $\sigma_B^{(3)}$ at $\mu=m_D$. There is no phenomenological estimate for them at this stage.

\begin{table}[htbp]
\caption{Upper panel: Numerical results for $\lambda_B$ and $\sigma_B^{(1,2)}$ at $\mu=m_D$, which are obtained from the fits based on Eq.(\ref{eq:HQETLCDAsmallomegaexpansion}) up to the $N$-th order, and the fits based on Eq.(\ref{eq:HQETLCDAsmallomegaexpansion2}) up to the $K$-th order.  Central panel: Results of $\lambda_B$ and $\sigma_B^{(1)}$ at $\mu=1$GeV after RG evolution, which is consistent with the scale of theoretical results pesented in the lower panel. Lower panel: $\lambda_B$ and $\sigma_B^{(1)}$ from experiment and other theoretical approaches. 
}
  \renewcommand{\arraystretch}{2.0}
  \setlength{\tabcolsep}{2.5mm}
  \begin{tabular}{c c c c c }
    \hline\hline
    $\mu$ & & $\lambda_B$ (GeV) & $\sigma_B^{(1)}$ & $\sigma_B^{(2)}$ \\
    \hline
    & $N=1$ & 0.424(41) & 2.17(12) & 6.36(51) \\
    & $N=2$ & 0.428(42) & 2.16(10)  & 6.29(47) \\
    \multirow{2}{*}{$m_D$}& $N=3$ & 0.418(67) & 2.18(15) & 6.33(80) \\ 
    & $K=0$ & 0.421(33) & 2.17(8) & 6.35(32) \\
    & $K=1$ & 0.424(35) & 2.18(8) & 6.36(32) \\
    & $K=2$ & 0.396(45) & 2.12(12) & 6.27(41) \\
    \hline
    & $N=1$ & 0.380(37)  & 1.66(9) &  \\
    & $N=2$ & 0.384(38) & 1.65(8) &  \\
   \multirow{2}{*}{1GeV} & $N=3$ & 0.374(60) & 1.66(12) &  \\ 
    & $K=0$ & 0.377(30) & 1.66(6) &   \\
    & $K=1$ & 0.380(31) & 1.67(6) &   \\
    & $K=2$ & 0.356(41) & 1.62(9) &   \\
    \hline  
    & Ref.\cite{Belle:2018jqd} & $>0.24$ & & \\
    & Ref.\cite{Khodjamirian:2020hob} & $0.383(153)$ & & \\
   \multirow{2}{*}{1GeV} & Ref.\cite{Lee:2005gza} & $0.48(11)$ & $1.6(2)$ & \\
    & Ref.\cite{Braun:2003wx} & $0.46(11)$ & $1.4(4)$ & \\
    & Ref.\cite{Grozin:1996pq} & $0.35(15)$ & & \\
    & Ref.~\cite{Gao:2019lta} & $0.343^{+0.064}_{-0.079}$ & & \\
    & Ref.\cite{Mandal:2023lhp} & $0.338(68)$ & & \\
    \hline
  \end{tabular} 
  \label{tab:model_independent_moments}
\end{table}
  
 {For a comparison, we present the results of inverse and inverse-logarithmic moments from experimental data and other theoretical approaches in the lower panel of Table \ref{tab:model_independent_moments}.}
Using the measurement of $B\to \gamma\ell\nu$, the Belle collaboration has provided a lower bound of $\lambda_B > 0.24$ GeV at a $90\%$ confidence level \cite{Belle:2018jqd}. Further constraints on $\lambda_B$~\cite{Khodjamirian:2020hob,Braun:2003wx,Grozin:1996pq} and $\sigma_B^{(1)}$~\cite{Braun:2003wx} are derived from QCD sum rules. 
Additionally, Ref.~\cite{Lee:2005gza} offers results for $\lambda_B$
and $\sigma_B^{(1)}$  using a realistic model description of the B-meson LCDA. 
In Ref.\cite{Mandal:2023lhp}, $\lambda_B$ is extracted by fitting the  lattice results of the $B\to K$ form factors.

Comparing the results presented in Table \ref{tab:model_independent_moments}, we observe that our findings for $\lambda_B$  are in agreement with those derived from experimental data and other theoretical methods. While the results for $\sigma_B^{(1)}$  are only available from certain theoretical studies, it is notable that our results are still in approximate agreement with them after performing the scale evolution.

By integrating all results for $\lambda_B$ and $\sigma_B^{(1,2)}$ from various parameterizations and expansions to different orders, as detailed in Table \ref{tab:model_independent_moments}, we obtain the corresponding constraints:
\begin{align}
    &\lambda_B(\mu=m_D)  = 0.420(71)~ \mathrm{GeV}, \nn\\
    &\sigma_B^{(1)}(\mu=m_D)  = 2.17(16), \nn\\
    &\sigma_B^{(2)}(\mu=m_D)  = 6.33(80), \nn\\
	&\lambda_B(\mu=1~\mathrm{GeV}) =0.376(63)~ \mathrm{GeV}, \nn\\
	&\sigma_B^{(1)}(\mu=1~\mathrm{GeV})  = 1.66(13).
\end{align}

\subsection{Impact on $B\to V$ form factors}

In Ref.~\cite{Gao:2019lta}, the $B\to V$ form factors have been updated  in LCSRs using $B$-meson light-cone distribution amplitudes. In addition to the next-to-leading order QCD corrections, the light-quark mass effect for the local soft-collinear effective theory form factors   is also computed from the LCSR method. Furthermore, the subleading power corrections to $B\to V$ form factors from the higher-twist B-meson light-cone distribution amplitudes are also computed with the same method at tree level up to the twist-six accuracy.

Instead of the ordinary form factors, the authors of Ref.~\cite{Gao:2019lta} have calculated the following form factors:
\begin{align}
{\cal V}(q^2) =& \frac{m_B}{m_B+m_V}V(q^2), \nonumber\\
{\cal A}_0(q^2)=& \frac{m_V}{E_V}A_0(q^2) 
, \nonumber\\
{\cal A}_1(q^2)=& \frac{m_B+m_V}{2E_V}A_1(q^2)  
,\nonumber\\
{\cal A}_{12}(q^2)=&\frac{m_B+m_V}{2E_V}A_1(q^2) -\frac{m_B-m_V}{m_B}A_2(q^2),
\nonumber\\
{\cal T}_1(q^2)=& T_1(q^2), \nonumber\\
{\cal T}_2(q^2)=& \frac{m_B}{2E_V} T_2(q^2) 
,\nonumber\\
{\cal T}_{23}(q^2)=& {\cal T}_2(q^2)-T_3(q^2),
\end{align}
where $m_V$ and $E_V$ denote the mass and energy for the vector meson, respectively. $m_B$ is the $B$ meson mass.  
At $q^2=0$, ${\cal A}_0(0)={\cal A}_{12}(0)$ and ${\cal T}_1(0)= {\cal T}_2(0)$. 

\begin{figure}[htbp]
\centering
\includegraphics[scale=0.7]{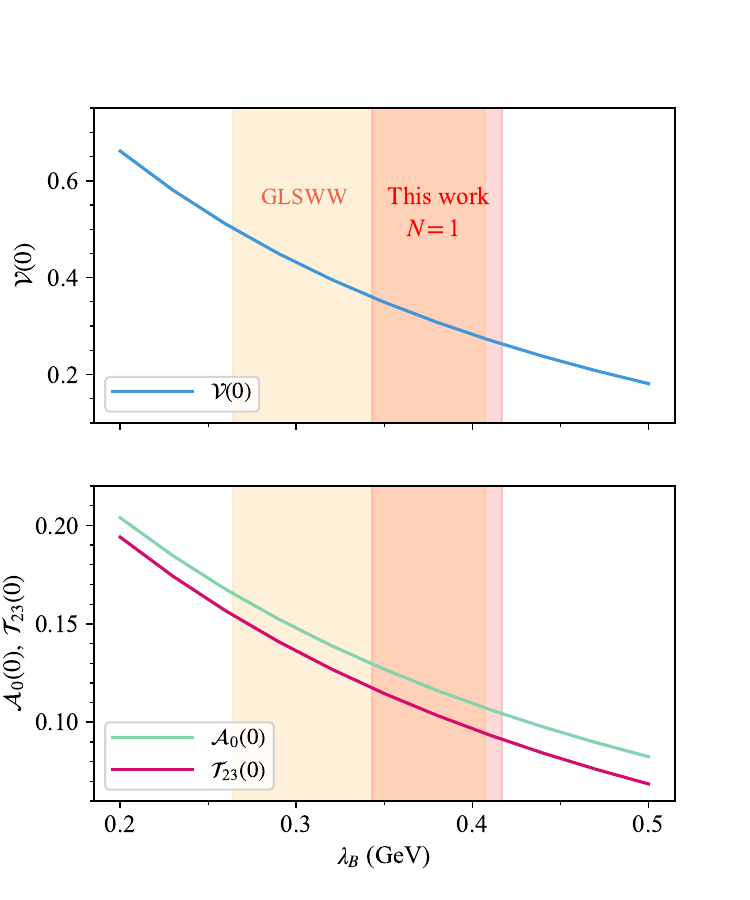}
\caption{{The dependence of the $B\to K^*$ form factors on the inverse moment of heavy meson LCDAs in the exponential model based on the results in Ref.~\cite{Gao:2019lta}.  The band labeled as “GLSWW” represents the results obtained using the inverse moment directly from Ref.~\cite{Gao:2019lta}, as shown in Eq.~\ref{eq:inverse_GLSWW}. On the other hand, the band labeled as “This work” corresponds to the inverse moment calculated in this paper. }}
  \label{fig:B_Kstar_formfactors}
\end{figure}

Using these analytical results,  we present the dependence of the $B\to K^*$ form factors on the first inverse moment in Fig.~\ref{fig:B_Kstar_formfactors}. {In alignment with the conventions of Ref.~\cite{Gao:2019lta}, we have adopted the results for $N=1$. Given that ${\cal V}$, ${\cal T}_1$, and ${\cal A}_1$ exhibit very similar numerical values, we have chosen to only display the results for ${\cal V}$ and ${\cal A}_0, {\cal T}_{23}$ at $q^2=0$.}
The band labeled as “GLSWW” represents the results obtained using the inverse moment directly from Ref.~\cite{Gao:2019lta}, as shown in Eq.~\ref{eq:inverse_GLSWW}. On the other hand, the band labeled as “This work” corresponds to the inverse moment calculated in this paper.
It is important to mention that in Ref.~\cite{Gao:2019lta}, the exponential model for the $B$ meson LCDAs utilizes the first inverse moment  as:
\begin{eqnarray}
\lambda_B = (0.343^{+0.064}_{-0.079}){\rm GeV}.  \label{eq:inverse_GLSWW}
\end{eqnarray}
This value is obtained by matching the results for $B\to \rho$ form factors onto the QCD LCSRs using light meson LCDAs.

From Fig.~\ref{fig:B_Kstar_formfactors}, we can see that the form factors computed using our inverse moment are in agreement with those in Ref.~\cite{Gao:2019lta}, although it should be stressed that systematic uncertainties are not included in this paper. From the results for ${\cal V}^{B\to K^*}(0)$ in Eq.~\ref{eq:B-LCSR_formfacor}, one can observe that a precise determination of $\lambda_B$  can lead to an accurate prediction. More importantly, our  results  from first-principles of QCD help to remove the primary uncertainties arising from the model parametrizations of heavy meson LCDAs.  This indicates the potential for a more precise analysis of  form factors  and accordingly physical observables, like the determination of $|V_{ub}|$. 

\section{Summary}

In this study, a unique framework for determining HQET LCDAs through a sequential effective theory approach is introduced and elaborated upon. The theoretical underpinnings of the framework have been discussed, providing a comprehensive overview of the methodology.  We have then delved into the intricacies of lattice QCD simulations, emphasizing the importance of analyzing dispersion relations and implementing fitting strategies to extract meaningful results. The calculations of quasi DAs, QCD LCDAs, and HQET LCDAs are showcased, offering valuable insights into the structure and behavior of heavy quark systems. Discussions focuses on the potential phenomenological implications of the findings, highlighting the implications for future research in the field of high-energy physics.  The potential phenomenological implications of the results are explored, offering insights into how these findings could influence our comprehension of the dynamics of strong and weak  interactions.

It should be emphasized that the current results are based on the simulation of quasi DAs with a single lattice spacing and perturbative calculation at leading power, where the predictions are likely to be affected by various systematic uncertainties from both lattice and analytical sides.   In the future, there are several areas of research that can be explored to further advance our understanding of heavy quark systems through the framework of HQET LCDAs using a sequential effective theory approach. Here are some potential avenues for future investigation. 
\begin{itemize}
\item  Building upon the framework of heavy quark spin symmetry, delving into the properties and behaviors of heavy vector mesons within the context of HQET LCDAs could yield valuable insights into the dynamics of these systems~\cite{Deng:2024dkd}. Exploring their quasi-distribution amplitudes and QCD local distribution amplitudes can deepen our comprehension of the structure of heavy quark systems.

\item  Utilizing smaller lattice spacings and larger momentum values in lattice QCD simulations can help improve the accuracy and precision of our calculations. Together with the investigation of lattice artifacts such as operator mixing effects,  they can lead to more reliable results and a better understanding of the behavior of heavy quark systems at different energy scales.

\item   It is conjectured that the operator mixing effects can be estimated by  studying the momentum dependence of the difference between the matrix elements with $\Gamma=\gamma^z\gamma_5$  and $\Gamma=\gamma^t\gamma_5$. Since the power corrections are expected to be momentum dependent and more importantly suppressed in the large momentum limit, the difference between the matrix elements after taking the large momentum limit is mostly attributed to the operator mixing effects, particularly from the $z^\mu$ term. This provides an estimate of the operator mixing effects.

Furthermore, since the zero-momentum matrix element with $\Gamma=\gamma^z\gamma_5$ has no contribution from the $P^\mu$ term, the non-zero value of this matrix element also indicates the magnitude of the operator mixing effects. It is essential to calculate this matrix element in the future, especially in comparison to the one with $\Gamma=\gamma^t\gamma_5$.

\item   Exploring a wider range of masses for the heavy quark and studying their dependence on various parameters can provide a more comprehensive picture of the behavior of heavy quark systems~\cite{Wang:2024wwa}. This can help elucidate the effects of mass on the structure and dynamics of these systems. An analysis using different heavy quark mass is under progress. 

\item The perturbative results  employed in this study are at the next-to-leading order in $\alpha_s$. Accounting for higher-order or instance two-loop  corrections has the potential to substantially diminish the associated uncertainties. This involves the perturbative calculation of two-loop matching kernels in two steps, and the  HQET LCDAs in the endpoint region.

\item Our current focus is mainly on $\varphi_B^+(\omega)$, while investigating the function $\varphi_B^-(\omega)$, which describes another leading-power  distribution amplitudes of heavy quark systems, can offer valuable insights into the internal structure and properties of these systems. Understanding the behavior of $\varphi^{-}(\omega)$ can also help refine our theoretical framework and improve our theoretical predictions.

\item As an intermediate step, we have provided results for the QCD LCDAs for the $D$  meson which are also of  great importance.  The QCD LCDAs are expected to be normalized to 1. Upon matching the quasi-distribution amplitudes to the QCD LCDAs, we initially find that the normalization condition can be met. However, after undergoing  renormalization group  evolution, we observe a universal decrease in the distributions, leading to a normalization constant smaller than 1. In our current manuscript, we have adopted a similar approach as referenced in Ref.~\cite{Beneke:2023nmj},  where the QCD LCDAs were also not properly normalized, albeit for different reasons.

A way to understand this behavior is to directly compute some lowest moments of the QCD LCDAs for charmed mesons. This complementary approach allows a direct comparison  of the results~\cite{Ji:2022ezo}, and perhaps some insights can be obtained in this way.

\end{itemize}
By exploring these areas of research in the future, we can further enhance our understanding of heavy quark systems and advance the field of high-energy physics.

\section*{Acknowledgement}

We thank Yu-Ming Wang,  Yan-Bing Wei, Xiaonu Xiong and Yong Zhao for  valuable  discussions, and particularly Yuming Wang for providing the error budget for $B\to \pi$ form factors and his code to calculate the $B\to V$ form factors in $B$ meson LCSRs.  We thank Thorsten Feldmann for drawing our attention to the interesting parametrization of heavy meson LCDAs. We thank Yue-Long Shen and  Gael Finauri for discussing  the importance of QCD LCDAs for the $D$ meson.   We thank the CLQCD collaborations for providing us the gauge configurations with dynamical fermions~\cite{Hu:2023jet}, which are generated on the HPC Cluster of ITP-CAS, the Southern Nuclear Science Computing Center(SNSC), the Siyuan-1 cluster supported by the Center for High Performance Computing at Shanghai Jiao Tong University and the Dongjiang Yuan Intelligent Computing Center. This work is supported in part by Natural Science Foundation of China under grant No. 12125503, 12335003, 12375069, 12105247, 12275277, 12293060, 12293062, 12047503, 12435004, 12475098, 12375080, 11975051, 12205106, and by the National Key Research and Development Program of China (2023YFA1606000).  A.S., W.W, Y.Y and J.H.Z are supported by a NSFC-DFG joint grant under grant No. 12061131006 and SCHA 458/22. Y.Y is supposed by the Strategic Priority Research Program of Chinese Academy of Sciences, Grant No.\ XDB34030303 and YSBR-101.  J.H.Z is supported by CUHK-Shenzhen under grant No. UDF01002851. Q.A.Z is supported by the Fundamental Research Funds for the Central Universities. J.H is supported by Guang-dong Major Project of Basic and Applied Basic Research No. 2020B0301030008.
The computations in this paper were run on the Siyuan-1 cluster supported by the Center for High Performance Computing at Shanghai Jiao Tong University, and Advanced Computing East China Sub-center. The LQCD simulations were performed using the Chroma software suite~\cite{Edwards:2004sx} and QUDA~\cite{Clark:2009wm,Babich:2011np,Clark:2016rdz} through HIP programming model~\cite{Bi:2020wpt}. This work was partially supported by SJTU Kunpeng \& Ascend Center of Excellence.


\end{document}